\documentclass[11pt]{article}
\usepackage{amsfonts,slashed}
\usepackage{float}
\usepackage{latexsym}
\usepackage{amsfonts}
\usepackage{amsmath}
\usepackage{epsfig}
\usepackage{latexsym,amssymb}
\usepackage{amsmath,amssymb,amsthm}
\usepackage{hyperref}
\usepackage{graphicx,tikz}
\usepackage{multirow}
\usepackage{textcomp}
\usepackage{cite}

\usepackage[shadow,textwidth=2.7cm]{todonotes}
\usepackage{ifthen}
\reversemarginpar
\newcounter{todocounter}

\colorlet{tdcolor}{red!30!white}

\newcommand{\todoinline}[2][]{
  \ifthenelse { \equal {#1} {} }
    { \def\temp {#2} }
    { \def\temp {#1} }
  \refstepcounter{todocounter}\todo[color=tdcolor,inline,caption={\textbf{\thetodocounter. } \temp}]{\textbf{\thetodocounter. } #2}{}}

\DeclareFontFamily{OMS}{rsfs}{\skewchar\font'60}
\DeclareFontShape{OMS}{rsfs}{m}{n}{<-5>rsfs5 <5-7>rsfs7 <7->rsfs10 }{}
\DeclareSymbolFont{rsfs}{OMS}{rsfs}{m}{n}
\DeclareSymbolFontAlphabet{\Scr}{rsfs}

\numberwithin{equation}{section}
\def\be{\begin{equation}}
\def\ee{\end{equation}}
\def\ba{\begin{array}}
\def\ea{\end{array}}

\newcommand{\bea}{\begin{eqnarray}}
\newcommand{\eea}{\end{eqnarray}}

\def\={~=~}
\def\*{{}^*}

\def\U{\mathrm{U}}
\def\SU{\mathrm{SU}}

\newcommand{\uA}{{\underline{A}}}
\newcommand{\uB}{{\underline{B}}}
\newcommand{\uC}{{\underline{C}}}
\newcommand{\uD}{{\underline{D}}}
\newcommand{\uE}{{\underline{E}}}
\newcommand{\uF}{{\underline{F}}}

\newcommand{\AI}{\left(A^{-1}(\eta)\right)}

\def\={~=~}
\def\*{{}^*}

\def\U{\mathrm{U}}
\def\SU{\mathrm{SU}}

\newcommand{\uphi}{\Phi}

\begin{document}
\begin{titlepage}
\vfill
\begin{flushright}
HU-EP-21/08
\end{flushright}

\vfill
\begin{center}
{\LARGE \bf Global Properties of the Conformal Manifold \\[1ex]
for S--Fold Backgrounds}\\[1cm]

{\large\bf Alfredo Giambrone\,$^{a,b}{\!}$
\footnote{\tt alfredo.giambrone@polito.it}, Emanuel Malek\,${}^{c}{\!}$
\footnote{\tt emanuel.malek@physik.hu-berlin.de}, Henning Samtleben\,${}^{d}{\!}$
\footnote{\tt henning.samtleben@ens-lyon.fr} \\[1ex]
and Mario Trigiante\,${}^{a,b}{\!}$
\footnote{\tt mario.trigiante@polito.it}} \vskip .8cm

{\it ${}^a$ Department of Applied Science and Technology, Politecnico di Torino, \\
Corso Duca degli Abruzzi 24,
I-10129 Torino, Italy}\\ \ \\
{\it  ${}^b$ INFN -- Sezione di Torino,\\
Via P. Giuria 1, 10125 Torino, Italy}\\ \ \\
{\it ${}^c$ Institut f\"{u}r Physik, Humboldt-Universit\"{a}t zu Berlin,\\
	IRIS Geb\"{a}ude, Zum Gro{\ss}en Windkanal 2, 12489 Berlin, Germany}\\ \ \\
{\it ${}^d$ Univ Lyon, Ens de Lyon, Univ Claude Bernard, CNRS,\\
Laboratoire de Physique, F-69342 Lyon, France}

\end{center}
\vfill

\begin{center}
\textbf{Abstract}

\end{center}
\begin{quote}
We study a one-parameter family of $\mathcal{N}=2$ anti-de Sitter vacua with ${\rm U}(1)^2$ symmetry of gauged four-dimensional maximal supergravity, with dyonic gauge group $[{\rm SO}(6)\times {\rm SO}(1,1)]\ltimes \mathbb{R}^{12}$. These backgrounds are known to correspond to Type IIB S-fold solutions with internal manifold of topology $S^1\times S^5$. The family of AdS$_4$ vacua is parametrized by a modulus $\chi$. Although $\chi$ appears non-compact in the four-dimensional supergravity, we show that this is just an artefact of the four-dimensional description. We give the 10-dimensional geometric interpretation of the modulus and show that it actually has periodicity of $\frac{2\pi}{T}$, which is the inverse radius of $S^1$. We deduce this by providing the explicit $D=10$ uplift of the family of vacua as well as computing the entire modulus-dependent Kaluza-Klein spectrum as a function of $\chi$. At the special values $\chi=0$ and $\chi=\frac{\pi}{T}$, the symmetry enhances according to ${\rm U}(1)^2\rightarrow{\rm U}(2)$, giving rise however to inequivalent Kaluza-Klein spectra. At $\chi=\frac{\pi}{T}$, this realizes a bosonic version of the ``space invaders'' scenario with additional massless vector fields arising from formerly massive fields at higher Kaluza-Klein levels.

\end{quote}
\vfill
\setcounter{footnote}{0}

\end{titlepage}
\tableofcontents \noindent {}
\newpage
\section{Introduction}
The duality covariant formulation of gauged supergravities in various dimensions \cite{deWit:2005ub,deWit:2007kvg,deWit:2005hv,deWit:2008ta} (see \cite{Samtleben:2008pe,Trigiante:2016mnt} for reviews), based on the notion of the \emph{embedding tensor} \cite{Cordaro:1998tx,Nicolai:2000sc,deWit:2002vt}, has provided a valuable tool for discovering new superstring/M-theory compactifications and their duality connections. A consistent truncation of the low-lying  modes of superstring/M-theory, in certain compactifications to $D$-dimensions, is captured by the an effective $D$-dimensional extended supergravity theory whose Lagrangian typically exhibits characteristic minimal couplings, associated with a gauge group $\mathcal{G}$, Yukawa terms and a scalar potential. All these features of the effective low-energy description depend on general characteristics of the higher-dimensional background, such as the geometry
of the internal manifold $M_{{\rm int}}$ and various kinds of fluxes which are present in  the solution. Interestingly they can be all encoded in a single object called the embedding tensor. This tensor is formally covariant with respect to the on-shell global symmetry group $G$ (sometimes referred to as the \emph{duality group}) of the corresponding \emph{ungauged} version of the same theory, namely of a $D$-dimensional supergravity with the same amount of supersymmetry and field content but no minimal couplings. Although the presence of minimal couplings typically breaks $G$, formal $G$-invariance of the field equations and the Bianchi identities are preserved, provided the embedding tensor is transformed together with all the other fields.
\par
As far as maximal supergravities are concerned, the on-shell global symmetry group is of exceptional type $G={\rm E}_{11-D}$.  In these cases a direct relation between certain gauged models and superstring/M-theory can be established within the framework of \emph{Exceptional Field Theory} (ExFT) \cite{Hohm:2013vpa,Hohm:2013uia,Hohm:2014fxa,Baguet:2015xha}. The latter provides a manifestly ${\rm E}_{11-D}$-covariant description of $11$--dimensional and Type--II supergravities and shows how to embed certain $D$-dimensional gauged supergravities within the higher-dimensional ones, as consistent truncations, through  a generalized Scherk--Schwarz ansatz \cite{Hohm:2014qga}. Recently, this framework has also proven to be very useful in performing Kaluza--Klein spectrometry for those compactifications fitting into the generalized Scherk--Schwarz ansatz \cite{Malek:2019eaz,Malek:2020yue}. As a key simplification, the construction only relies on the scalar harmonics, corresponding to the maximally symmetric point of the lower dimensional supergravity. \par
Over the last ten years or so, new classes of gaugings were found in four-dimensions, which involved, in a standard symplectic frame, magnetic components of the embedding tensor (\emph{dyonic gaugings}) \cite{DallAgata:2011aa,DallAgata:2012mfj,DallAgata:2014tph}. While some of these models are constructed by gauging a same simple gauge group of the form ${\rm SO}(p,q)$ in different frames,  others involve non-semisimple gauge groups and have the general form $[{\rm SO}(p,q)\times {\rm SO}(p',q')]\ltimes N$, with $N$ being a subgroup generated by nilpotent generators. The dyonic nature of the latter gaugings (i.e.\ the non-vanishing magnetic components of the embedding tensor) is encoded in a deformation parameter $c$ which
if non-vanishing
can always be set to a fixed value by field redefinitions, e.g. $c=1$. All these gaugings generalize their electric simple and semi-simple counterparts \cite{deWit:1982bul,Hull:1988jw} (the non-semisimple gaugings, for $c=0$, reduce to the electric ${\rm CSO}(p,q,r)$ gaugings).

As was shown in a series of works, the non-semisimple dyonic gaugings can be embedded in Type~II supergravity. For example, the dyonic ${\rm ISO}(7)$-model was shown to be a consistent truncation of massive Type IIA supergravity \cite{Romans:1985tz} on a background of the form ${\rm AdS}_4\times S^6$ \cite{Guarino:2015jca,Guarino:2015vca,Guarino:2015qaa,Guarino:2019snw,Guarino:2020jwv,Guarino:2020flh}. The general embedding of the models featuring non-semisimple dyonic gaugings within Type II supergravities was derived, employing the ExFT framework, in \cite{Inverso:2016eet}.\par
Here we are interested in the four-dimensional maximal supergravity with dyonic gauging
\begin{equation}
\mathcal{G}=[{\rm SO}(6)\times {\rm SO}(1,1)]\ltimes \mathbb{R}^{12}\,,
\end{equation}
which features AdS$_4$ vacua with $\mathcal{N}=0,\,1,\,2$ and $4$ supersymmetries \cite{Gallerati:2014xra,Inverso:2016eet,Guarino:2019oct,Guarino:2020gfe}. Some of these were lifted to Type IIB S-folds of Janus solutions, which have a spacetime geometry of the form AdS$_4\times S^1\times S^5$, with $S^5$ being a deformed five-sphere. These backgrounds are characterized by a monodromy $\mathfrak{M}_{S^1}$ around the non-contractible $S^1$ with radius $\frac{T}{2\pi}$, with $\mathfrak{M}_{S^1}$ a hyperbolic element of the ${\rm SL}(2,\mathbb{Z})_{\rm IIB}$ duality group. In other words, these solutions feature different local geometric descriptions patched together by a non-perturbative Type IIB S-duality transformation. They can also be constructed as suitable quotients of Janus-like solutions in Type IIB \cite{Bak:2003jk,DHoker:2006vfr}.

The $\mathcal{N}=4$ vacuum with ${\rm SO}(4)$ residual gauge symmetry was found in \cite{Gallerati:2014xra} and uplifted to Type IIB theory in \cite{Inverso:2016eet}. The $\mathcal{N}=0,\,1$ vacua were discovered in \cite{Guarino:2019oct}. The $\mathcal{N}=0$ vacuum with symmetry ${\rm SU}(4)$ and the $\mathcal{N}=1$ one with symmetry ${\rm SU}(3)$ were uplifted, in the same work, to ten-dimensional S-folds of type IIB. In \cite{Guarino:2020gfe}, a new family of $\mathcal{N}=2$ ${\rm U}(1)^2$ symmetric vacua was found. The vacua of this family are labeled by a continuous, non-compact parameter $\chi$.\footnote{In fact this family of vacua will feature at least two moduli fields, as the conformal manifold ought to be complex. The supergravity moduli fields are expected to be a subset of the four scalar massless modes found in \cite{Guarino:2020gfe}. Recently a 2-parameter extension of the $\mathcal{N}=2$ vacua studied here was constructed in \cite{Arav:2021gra,Bobev:2021yya}.} At $\chi=0$ the residual gauge symmetry is enhanced to ${\rm SU}(2)\times{\rm U}(1)$ and the type IIB uplift at this particular value was found in the same work.

The corresponding S-fold solutions are conjectured to be holographically dual to interface super-Yang Mills theories in $D=3$.
Interesting examples are given in \cite{Assel:2018vtq}, where a class of S-fold $\mathcal{N}=4$ AdS$_4\times K_6$ solutions with compact $K_6$ internal manifold is given. Following the authors, these solutions can be obtained as quotients of known non-compact ones, with the quotient defined by an ${\rm SL}(2,\mathbb{Z})_{{\rm IIB}}$ action on the latter. Furthermore, by translating this procedure on the corresponding $\mathcal{N}=4$ CFT$_3$ Janus-type theories \cite{DHoker:2006qeo}\cite{Gaiotto:2008sd}, they were able to find strong candidates for their SCFT$_3$ duals.\par
Let us now summarize and briefly discuss the results of the present paper.
By employing  the ExFT methods, we perform a Kaluza-Klein analysis on the ${\rm U}(1)^2$-symmetric $\mathcal{N}=2$ family of vacua found in \cite{Guarino:2020gfe}. We perform their uplift to Type IIB S-fold solutions of the whole 1-parameter  $\mathcal{N}=2$ family. In particular, we give $\chi$ a geometrical interpretation as a 10-dimensional metric modulus. We find that the dependence on $\chi$ of the type IIB solution can be interpreted as a global twist in the internal geometry, and, in particular, involving a squashed $S^3$ submanifold of the deformed $S^5$, which is fibered over $S^1$. This fibration involves a non-trivial twist of the points of $S^3$, as we move around $S^1$, which depends on $\chi$.

The way this occurs can be understood as follows. Let us denote by $\eta$ the compact $S^1$ coordinate, in the interval $[0,\,T)$, and by $\alpha,\beta\,\gamma$ the three angular coordinates of the deformed $S^3$. For $\chi=0$, a generic point of the latter manifold, is described by the $ {\rm SU}(2)$ group element $g(\alpha,\beta,\gamma)$. The metric of the internal manifold features an $ {\rm SU}(2) $ isometry group  acting from the left while the squashing of $S^3$ breaks the $ {\rm SU}(2)'$ isometry which, in the round $S^3$, would act from the right on the same element, to a ${\rm U}(1)' $ subgroup of it. For $\chi\neq 0$ the fibration of $S^3$ over $S^1$ is affected by redefining the $ {\rm SU}(2)$ element describing a point in $S^3$ as follows:
\begin{equation}
g(\alpha,\beta,\gamma)\,\rightarrow\,\,g(\alpha',\beta',\gamma')=h(\eta)\cdot g(\alpha,\beta,\gamma)\,\,,\,\,\,\,h(\eta)\equiv \left(\begin{matrix}\cos( \chi \eta) & \sin( \chi \eta)\cr -\sin( \chi \eta) & \cos( \chi \eta)\end{matrix}\right)\,.\label{twistchi}
\end{equation}
Locally this change can be reabsorbed in a reparametrization of $S^3\times S^1$ $\{\alpha,\beta,\gamma,\eta\}\,\rightarrow\,\{\alpha',\beta',\gamma',\eta'\} $, where $\eta'=\eta$ and $\alpha',\beta',\gamma'$ are defined by the matrix equation in (\ref{twistchi}).

In fact, as we shall prove, the $D=10$ S-fold solutions corresponding to the $\chi\neq 0$ vacua are locally related to the one associated with the $\chi= 0$ vacuum by the above  reparametrization, although globally different. In particular, $\chi$ only enters through the dependence of the fields on the point of the squashed $S^3$ and thus does not affect the axion-dilaton field.

Note that the matrix $h(\eta)$ in (\ref{twistchi}) induces a non-trivial monodromy $h(T)$ as $\eta\,\rightarrow \,\eta+T$ which breaks the $ {\rm SU}(2) $ internal isometry of the $\chi=0$ solution, to the ${\rm U}(1) $ subgroup commuting with $h(T)$. From the expression of $h(\eta)$ it follows that for $\chi=\frac{2\pi}{kT}$, with $k$ positive integer, the monodromy matrix $h(T)$, acting on $g(\alpha,\beta,\gamma)$ from the left, generates the $\mathbb{Z}_k$ cyclic group and, if $k=1, 2$, the $ {\rm SU}(2)$ isometry is unbroken.\footnote{For $k=2$ the $ {\rm SU}(2)$ group commutes with $\mathbb{Z}_2$, since $\mathbb{Z}_2$ is its center.}

We are also able to relate $\chi$ with a \emph{complex structure modulus} associated with the internal submanifold $S^3\times S^1$. Indeed, writing $S^3$ as a Hopf fibration of a circle over $S^2$ and combining the circular fibre with the external $S^1$ into a 2-torus $T^2$, the manifold $S^3\times S^1$ can be written as a toroidal fibration over $S^2$. We show that $\chi$ defines the real part of the modular parameter of the toroidal fiber $T^2$ and, due to the invariance of the complex structure of the torus under a Dehn twist, $\chi$ has period $\frac{2\pi}{T}$.

All these global properties of the $D=10$ background, associated with the $\chi$ parameter, cannot be seen from the four-dimensional supergravity perspective, but are apparent from the analysis of the Kaluza-Klein spectrum of these vacua, which we perform. At the special values $\chi = \frac{\pi m}{T}$, $m \in \mathbb{Z}$, two vectors in the full KK spectrum, but outside the supergravity truncation, become massless, thus enhancing ${\rm U}(1)^2$ to ${\rm SU}(2)\times {\rm U}(1)'$. This corresponds to a \emph{space invaders} scenario \cite{Duff:1986hr,Cesaro:2020piw}. Moreover, for vacua related by the shift $\chi \rightarrow \chi + \frac{2\pi}{T}$, the entire Kaluza-Klein spectrum is identical, although differently distributed over the $S^1\times S^5$ KK levels along the lines described in \cite{Dabholkar:2002sy}\footnote{More precisely, this shift in $\chi$ can be reabsorbed in a redefinition of the Kaluza-Klein level $n$ on $S^1.$}, while for $\chi=\frac{\pi}{T}$, the Kaluza-Klein spectrum differs. This is consistent with the fact, outlined above, that the internal $\mathbb{Z}_2$-monodromy generated by $h(T)$, for $\chi=\frac{\pi}{T}$, is non-trivial, while still commuting with ${\rm SU}(2)$.\par
As far as the dual $3$-dimensional theory is concerned, we can still rely on the constructions put forward in \cite{Bobev:2020fon}, building on \cite{Assel:2018vtq}. One of these possibilities involves the strong coupling regime of the  ${\rm T}[{\rm U}(N)]$ theory by Gaiotto-Witten \cite{Gaiotto:2008ak} in which the ${\rm U}(N)\times {\rm U}(N)$ global symmetry is gauged by a ${\rm U}(N)$ $\mathcal{N}=2$ vector multiplet, so as to preserve $\mathcal{N}=2$ supersymmetry in the IR limit. The parameter $\chi$ would parametrize a further exactly marginal deformation of this  $\mathcal{N}=2$ model, thus defining a direction in the conformal manifold of the dual theory. Our analysis, unveiling the compact nature of $\chi$, sheds some light on the global properties of the conformal manifold.\par
The paper is organized as follows. In section \ref{TGS}, after a brief description of the gauged four-dimensional supergravity under consideration, we review the general features of the 1-parameter family of $\mathcal{N}=2$ vacua and give the corresponding  supermultiplet structure of the supergravity fields. In section \ref{KKExFT}, we perform the Kaluza-Klein analysis on the same vacua giving the bosonic mass spectrum up to level 3 and general mass formulae. In section \ref{uplift}, using the ExFT approach, we uplift the family of $\mathcal{N}=2$ vacua to S-fold solutions in Type IIB supergravity, we elaborate on the geometric interpretation of $\chi$.
We conclude with a final discussion.

\section{The Gauged $D=4$ Supergravity}\label{TGS}
In this section we describe the general structure of the four-dimensional supergravity we shall be working in.\par
Maximal supergravity in four dimensions only describes a gravitational multiplet consisting of the graviton, $8$ gravitini, $28$ vector fields, $56$ spin-$1/2$ fields and $70$ scalars spanning the scalar manifold ${\rm E}_{7(7)}/{\rm SU}(8)$.
The on-shell global symmetry group of the ungauged model is ${\rm E}_{7(7)}$ which  acts as an electric-magnetic duality group on the 28 vector fields strengths and their magnetic duals. This duality action is defined by the symplectic $56$-dimensional representation of ${\rm E}_{7(7)}$.
We shall be working in the symplectic frame in which the off-shell global symmetry group is ${\rm SL}(8,\mathbb{R})\subset {\rm E}_{7(7)}$ (${\rm SL}(8,\mathbb{R})$-symplectic frame). If $A,B=1,\dots, 8$ label the fundamental 8-dimensional representation of this group, the 28 electric vector fields $A^{[AB]}_\mu$ and their magnetic  counterparts $A_{[AB] \, \mu}$,\footnote{They are required for a manifest duality covariant formulation.} are labeled by the antisymmetric couple $[AB]$. These fields are conveniently described by a symplectic 56-component vector $A^M_\mu$, $M=1,\dots, 56$, of the form: $A^M_\mu=(A^{[AB]}_\mu,\,A_{[AB]\mu})$. \footnote{For the ${\rm SL}(8,\mathbb{R})\subset {\rm E}_{7(7)}$ indices we use the notation that contraction over an antisymmetric couple $[AB]$ should be multiplied times a factor $1/2$: $V_M\,W^M=\frac{1}{2}\,(V_{[AB]}W^{[AB]}+V^{[AB]}W_{[AB]})$.}
The generators of ${\rm E}_{7(7)}$ consist of the ${\rm SL}(8,\mathbb{R})$ generators $t^A{}_B$ and generators $t^{ABCD}=t^{[ABCD]}$ in the representation ${\bf 70}$ of the same group.\par
The gauged theory with gauge group $\mathcal{G}$ is obtained by promoting $\mathcal{G}$, subgroup of ${\rm E}_{7(7)}$, from a global symmetry group to a local symmetry one according to a well-defined procedure which ensures the $\mathcal{N}=8$ supersymmetry of the resulting model. This procedure implies the introduction of additional terms in the Lagrangian, which include a scalar potential, and in the supersymmetry transformation laws of the fermion fields~\cite{deWit:1982bul,deWit:2007kvg}.\par

In the symplectic-covariant formulation of the gauging procedure, the gauge algebra  is described by a 56-component symplectic vector
of generators $X_M$, $M=1,\dots, 56$, each represented by a matrix $(X_{M})_N{}^P$ in the symplectic 56-dimensional representation of the ${\rm E}_{7(7)}$ generators:
\begin{equation}
 X_{M N}{}^P\mathbb{C}_{QP}= X_{M Q}{}^P\mathbb{C}_{NP}\,,
\end{equation}
where $\mathbb{C}_{NP}$ is the antisymmetric $56\times 56$ symplectic invariant matrix and $X_{M N}{}^P\equiv (X_{M})_N{}^P$ is formally an ${\rm E}_{7(7)}$-tensor encoding all information about the embedding of the gauge algebra within the global symmetry one. It is therefore also called the \emph{embedding tensor}. All the additional terms, required by the gauging procedure, in the Lagrangian (Yukawa terms and scalar potential) and in the supersymmetry transformation laws are expressed in terms of $X_{MN}{}^P$.

The gauge connection is defined as follows:
\begin{equation}
\Omega_{g\,\mu}\equiv g\,A^M_\mu\,X_M=\frac{g}{2}\,\left(A^{[AB]}_\mu\,X_{[AB]}+A_{{[AB]}\,\mu}\,X^{[AB]}\right)\,,
\end{equation}
where $g$ is the gauge coupling. Besides $A_{{[AB]}\,\mu}$, also a set of antisymmetric 2-forms $B_{a\,\mu\nu}$, $a=1,\dots, 133$, transforming in the adjoint representation of ${\rm E}_{7(7)}$, has to be introduced. This is a redundant description of the field content which is required when we gauge a group $\mathcal{G}$ using vector fields which are not electric in the symplectic frame of the original ungauged Lagrangian.

 A set of linear and quadratic constraints on $X_{MN}{}^P$ guarantee the consistency of the gauging and in particular the existence of a symplectic frame in which the vector fields gauging $\mathcal{G}$ are electric.

 The scalar potential in terms of the embedding tensor reads:
\begin{equation}
 V(\phi)=\frac{g^2}{672}\,{M}^{MN}\,\left( {M}^{PQ} {M}_{RS}\,X_{MP}{}^R\,X_{NQ}{}^S+ 7\,{\rm Tr}(X_M\,X_N)\right)\,.\label{Vgeneral}
\end{equation}
The symplectic, symmetric matrix ${M}_{MN}(\phi)$ is defined in terms of the coset representative $\mathcal{V}(\phi)_M{}^N$ of the scalar manifold in the representation ${\bf 56}$ of ${\rm E}_{7(7)}$ as follows:
 \begin{equation}
{M}_{MN}(\phi)\equiv \mathcal{V}(\phi)_M{}^{\underline{A}} \mathcal{V}(\phi)_N{}^{\underline{A}} \,\in \,\frac{{\rm E}_{7(7)}}{{\rm SU}(8)}\,,
 \end{equation}
 where here and in the following we denote by $\underline{A},\,\underline{B}=1,\dots, 56$ the ${\rm SU}(8)$ indices labeling the ${\bf 28}+\overline{{\bf 28}}$ representation and summation over $\underline{A}$, in the above formula, is understood. In (\ref{Vgeneral}) ${M}^{MN}$ describes the inverse matrix of ${M}_{MN}$. The gauging procedure requires $X_{NM}{}^P$ to transform in the ${\bf 912}$ representation of ${\rm E}_{7(7)}$ (linear constraint) and to satisfy quadratic constraints which express the invariance of it under the action of the gauge group.\par
 We shall consider the gauged model in which the gauge group has the form
 \cite{DallAgata:2011aa,Inverso:2016eet,Guarino:2019oct,Guarino:2020gfe}:
\begin{equation}
\mathcal{G}=[{\rm SO}(6)\times {\rm SO}(1,1)]\ltimes \mathbb{R}^{12}\,.
\end{equation}
In the ${\rm SL}(8,\mathbb{R})$-symplectic frame the embedding tensor $X_{MN}{}^P$ of the gauging reads:
\begin{equation}\begin{split}
X_{[AB],\,[CD]}{}^{[EF]}&=-X_{[AB]}{}^{[EF]}{}_{[CD]}=8\,\delta_{[A}^{[E}\theta_{B][C}\delta_{D]}^{F]}\,,\\
X^{[AB]}{}_{[CD]}{}^{[EF]}&=-X^{[AB]\,[EF]}{}_{[CD]}=8\,\delta^{[A}_{[C}\xi^{B][E} \delta^{F]}_{D]}\,,\label{XMNP}
\end{split}\end{equation}
where
\begin{equation}
\theta_{AB}={\rm diag}(1,1,1,1,1,0,0,1)\,\,,\,\,\,\,\xi^{AB}={\rm diag}(0,0,0,0,0,1,-1,0)\,.
\end{equation}
Note that the ``magnetic'' vectors $A_{[AB]\,\mu}$ are involved in the gauge connection. \par
In our discussion about this model, we shall follow, unless stated otherwise, the notations of \cite{Guarino:2020gfe}. The model features anti-de Sitter vacua with supersymmetry $\mathcal{N}=0,\,1,\,2$ and $4$. We shall focus below on the $\mathcal{N}=2$ class of vacua, compute the (bosonic) Kaluza-Klein spectrum on them and eventually provide their uplift to $D=10$.
Following \cite{Guarino:2020gfe}, the vacua we are interested in can all be described within a  $\mathbb{Z}_2^3$-invariant sector \cite{Aldazabal:2006up}\cite{Guarino:2019snw} which describes an $\mathcal{N}=1$  supergravity coupled to seven chiral multiplets with complex scalars  $z_i=-\chi_i+ { i}\, e^{-\varphi_i}$, $i:1,...,7$. The coset representative, in a suitable basis of the ${\rm E}_{7(7)}$ generators, is chosen to be:

\begin{equation}
  \mathcal{V}={\rm exp}\left(\sum^{7}_{i=1} \chi_i e_i \right) \cdot {\rm exp}\left(\sum^{7}_{i=1} \varphi_i h_i\right) \in \left[\frac{{\rm SL}(2)}{{\rm SO}(2)}\right]^7\subset \frac{{\rm E}_{7(7)}}{{\rm SU}(8)}\,,
\end{equation}
where the generators $h_i,\,e_i$ satisfy the relations $[h_i,e_j]=\delta_{ij}\,e_j$, $[e_i,\,(e_j)^t]=2\delta_{ij}\,h_i$. They are related to the generators  $g_{\chi_i} ,\, g_{\varphi_i}$ in \cite{Guarino:2020gfe} as follows: $h_i=g_{\varphi_i}/4,\,e_i=-12\,g_{\chi_i}$.

\subsection{The $\mathcal{N}=2$ Vacua}
We shall focus our discussion to the $\mathcal{N}=2$ vacua, defined by the following expectation values for the scalars $z_i$:

\begin{equation}
  z_1=-\overline{z}_3=-\chi + \frac{i}{\sqrt{2}}\,, \qquad z_2=z_4=z_6=i\,, \qquad z_5=z_7=\frac{1}{\sqrt{2}}(1+i)\,.
\end{equation}
This family of vacua is parametrized by a continuous parameter $\chi$. From a low energy $D=4$ supergravity perspective, this parameter takes values in $\mathbb{R}$. It describes an ${\rm SU(2)}\times{\rm U}(1)$-invariant vacuum only when $\chi=0$. In general, for $\chi\neq 0$, the vacuum posses a ${\rm U}(1)^2$ residual gauge symmetry. As we shall see, this picture is strongly modified when considering the whole KK spectra of these backgrounds, or, equivalently, the corresponding $D=10$ solution. This analysis will show that $\chi$ is in fact periodic of period $2\pi/T$.

\subsubsection{${\rm U}(1)^2$ symmetric vacua}
In this case, there must be a massless gravity multiplet. It contains two gravitini with $m^2=1$
(massless in the corrected sense,
where all masses are normalized in units of $1/L=\sqrt{-V_0/3}=g$) and one massless vector.
Hence, only one massless vector multiplet must be considered with the other vectors being massive. Furthermore, they must come in pairs with opposite $R$--charges in order to fit into $\mathfrak{u}(1)_R$ representations. The remaining fields live in pairs of matter multiplets. The spectrum is organized into the following OSp$(2|4)$ supermultiplets
\begin{equation}\begin{split}
  &A_1\bar{A}_1[1]^{(0)}_{2}
    \oplus \;L\bar{A}_1[\textstyle\frac{1}{2}]^{(1)}_{\frac{5}{2}}\oplus \;A_1\bar{L}[\textstyle\frac{1}{2}]^{(-1)}_{\frac{5}{2}}
   \; \oplus 4 \times L\bar{L}[\tfrac12]^{0}_{\frac12+\sqrt{2+\chi^2}} \;\oplus A_2\bar{A}_2[0]^{(0)}_{1}
  \\
  &    \;\oplus L\bar{B}_1[0]^{(2)}_{2}\;\oplus B_1\bar{L}[0]^{(-2)}_{2}
 \; \; \oplus 2\times L\bar{L}[0]^{(0)}_{\frac{1}{2}+\frac{1}{2} \sqrt{1+16\chi^2}}
 \; \; \oplus 2\times L\bar{L}[0]^{(0)}_{\frac12+\frac{1}{2}\sqrt{17}}
    \;.
\end{split}
\label{level00}
\end{equation}
We refer to appendix~\ref{supermult} for notation and details on these multiplets.

\subsubsection{${\rm SU}(2) \times {\rm U}(1)$ symmetric vaccum}
At the supergravity level, when $\chi=0$, some of the long multiplets in (\ref{level00})
reach the unitarity bound and the following branching rule applies
\begin{equation}\begin{split}
  L\bar{L}[0]^{(0)}_{\frac{1}{2}+\frac{1}{2} \sqrt{1+16\chi^2}}
 \;\;\stackrel{\chi \rightarrow 0}{\longrightarrow} \;\;
   A_2\bar{A}_2[0]^{(0)}_1 \oplus L\bar{B}_1[0]^{(2)}_{2}\oplus    B_1\bar{L}[0]^{(-2)}_{2}\;.
  \label{shorteningLL10}
\end{split}\end{equation}
The resulting shortened multiplets join their copies in (\ref{level00}) to combine into an ${\rm SU}(2)$ vector.
In particular, two massive vectors become massless and join into the gauge vectors of the enhanced ${\rm SU}(2)$ symmetry.

\section{Embedding the Model in ExFT}

In this section, we shall use the framework of ${\rm E}_{7(7)}$-exceptional field theory (ExFT)  \cite{Hohm:2013uia}
to uplift the one-parameter $\mathcal{N}=2$ family of vacua to $D=10$ backgrounds of Type IIB supergravity.
ExFT is a reformulation of 10-/11-dimensional supergravity, which unifies the metric and flux degrees of freedom
within a manifestly E$_{7(7)}$ covariant formulation.
Its bosonic field content
\begin{equation}
	\begin{split}
	\left\{g_{\mu\nu}, {\cal M}_{MN},{\cal A}_\mu{}^M \right\}	\,, \quad&
	\mu=0, \dots,3\,,\\
	&{}
	M=1, \dots, 56
	\,,
	\label{fields}
	\end{split}
\end{equation}
contains an external and an internal metric $g_{\mu\nu}$, ${\cal M}_{MN}$, respectively,
with the latter parametrizing the coset space E$_{7(7)}/{\rm SU}(8)$,
together with vector fields, ${\cal A}_\mu{}^M$, transforming in the ${\bf 56}$ of the group E$_{7(7)}$.

${\rm E}_{7(7)}$-ExFT is defined on an extended spacetime spanned by the four-dimensional coordinates $x^\mu$, $\mu=0,1,2,3$, and 56 internal ones $Y^M$, $M=1,\dots, 56$, in the representation ${\bf 56}$ of ${\rm E}_{7(7)}$, subject to section constraints which
are satisfied if the fields are restricted to the original $D=11$ or IIB coordinates. The IIB diffeomorphisms and gauge symmetries combine into generalized diffeomorphisms on this extended spacetime. In order to perform the uplift of AdS$_4$ vacua, we only need the fields $g_{\mu\nu}(x,Y)$ and the generalized metric $\mathcal{M}_{MN}(x,Y)$ of the theory, the vector and tensor fields being consistently set to zero in this background. These fields are related to their counterparts $g_{\mu\nu}(x),\,M_{MN}(\phi(x))$ of the four-dimensional supergravity described in section \ref{TGS}, using the generalized Scherk-Schwarz ansatz \cite{Hohm:2014qga}:
\begin{equation}\begin{split}
g_{\mu\nu}(x,Y)&=\rho(Y)^{-2}\,g_{\mu\nu}(x)\,,\\
\mathcal{M}_{MN}(x,Y)&= U_{M}{}^{{K}}(Y)\,U_{N}{}^{{L}}(Y)\,M_{{KL}}(\phi(x))\,,\label{gMExFT}
\end{split}\end{equation}
where the twist matrix $U_{M}{}^N(Y)$ is associated with the gauging of the lower dimensional theory and defines the embedding of the  latter within the ExFT. The relationship between $U_{M}{}^N(Y)$ and the constant embedding tensor $X_{MN}{}^P$ (\ref{XMNP}) in the four-dimensional theory is:
\begin{equation}
\left.U^{-1}{}_M{}^RU^{-1}{}_N{}^Q\partial_R\,U_Q{}^P\right\vert_{{\bf 912}}=\frac{\rho}{7}\,X_{MN}{}^P\,,
\end{equation}
with the scalar function $\rho=\rho(Y)$ from (\ref{gMExFT}). Equivalently, this condition is expressed as~\cite{Lee:2014mla}
\begin{equation}
{\cal L}_{{\cal U}_M} {\cal U}_N = X_{MN}{}^P\,{\cal U}_P
\,,
\label{LUUXU}
\end{equation}
via the action of generalized diffeomorphisms, where the $\,{\cal U}_M$ denote the generalized
(56-dimensional) vectors
\begin{equation}
(\,{\cal U}_M)^N = \rho^{-1} (U^{-1})_M{}^N
\,,
\label{calU}
\end{equation}
given by the columns of the inverse twist matrix.

In our case, the twist matrix $U_{M}{}^N(Y)$ is an element of the ${\rm SL}(8,\mathbb{R})$ subgroup of ${\rm E}_{7(7)}$ and, in the symplectic basis of the ${\bf 56}$ representation in which ${\rm SL}(8,\mathbb{R})$ is diagonally embedded, reads:
\begin{equation}
U_{M}{}^N(Y)=\left(\begin{matrix}U_{[AB]}{}^{[CD]}(Y) & {\bf 0}\cr {\bf 0} & U^{[EF]}{}_{[GH]}(Y)=U^{-1}_{[GH]}{}^{[EF]}(Y)\end{matrix}\right)\,,
\label{twist_exp}
\end{equation}
where the $28\times 28$ matrix $U_{[AB]}{}^{[CD]}(Y)$ is expressed in terms of the $8\times 8$ one $U_{A}{}^{B}(Y)$, which describes the same ${\rm SL}(8,\mathbb{R})$ element in the representation ${\bf 8}$, as follows:
\begin{equation}
U_{[AB]}{}^{[CD]}=2\,U_{[A}{}^{[C}\,U_{B]}{}^{D]}\,.
\end{equation}

To embed Type IIB supergravity in ExFT we need the branching of the relevant ${\rm E}_{7(7)}$ representations with respect to the subgroup ${\rm SL}(6,\mathbb{R})\times {\rm SL}(2,\mathbb{R})_{{\rm IIB}}\times {\rm SO}(1,1)$, where ${\rm SL}(2,\mathbb{R})_{{\rm IIB}}$ is the global symmetry group of the Type-IIB:
\begin{equation}
{\bf 56}\,\rightarrow\,\,({\bf 6}',{\bf 1})_{-2}+({\bf 6},{\bf 2})_{-1}+({\bf 20},{\bf 1})_0+({\bf 6}',{\bf 2})_{+1}+({\bf 6},{\bf 1})_{+2}\,,
\end{equation}
the subscript being the ${\rm SO}(1,1)$-grading.
Correspondingly $Y^M$ splits as follows:
\begin{equation}
Y^M\,\rightarrow\,\,y^m\,,\,\,\,y_{\alpha m}\,,\,\,\,y_{mnp}\,,\,\,\,y^{\alpha m}\,,\,\,\,y_m\,,\label{Ydec}
\end{equation}
where $m,n,p=1,\dots, 6$  and $\alpha=1,2$ labels the components of an ${\rm SL}(2,\mathbb{R})_{{\rm IIB}}$ doublet. Restricting the ExFT fields to the $y^m$ coordinates only, the section constraints are satisfied and the field equations of ExFT reduce to those of Type IIB supergravity.
To identify the above components of $Y^M$ with the components of the same vector in the basis $Y^{[AB]},\,Y_{[AB]}$ it is necessary to further split the  ${\rm SL}(6,\mathbb{R})$ representations with respect to its ${\rm SL}(5,\mathbb{R})\times {\rm SO}(1,1)$ subgroup, so that $y^i$, $i=1,\dots, 5$, are identified with $Y^{[i8]}$ while $y^6$, to be denoted by $\tilde{y}$, is identified with $Y_{[67]}$, and we can write $(y^m)=(y^i,\,\tilde{y})$.  We refer to \cite{Inverso:2016eet} and \cite{Guarino:2019oct,Guarino:2020gfe} for the detailed correspondence between the quantities in the decomposition (\ref{Ydec}) and the components $Y^{[AB]},\,Y_{[AB]}$.\footnote{As opposed to the notations used in \cite{Guarino:2020gfe}, here we label by an upper (or lower) index $m$ a vector transforming in the ${\bf 6}'$  (or ${\bf 6}$) representation of ${\rm SL}(6,\mathbb{R})$.}
The explicit form of (the inverse of) $U_{A}{}^{B}(y^m)$ is given in \cite{Inverso:2016eet}.
\par

To express the components of the matrix $\mathcal{M}_{MN}(x,y)$, $y\equiv (y^m)$, in terms of $D=10$ fields we further need the decomposition of the
${\bf 133}$ of ${\rm E}_{7(7)}$, which branches as follows
\begin{equation}
{\bf 133} \rightarrow ({\bf 1},{\bf 2})_{+3}+({\bf 15}',{\bf 1})_{+2}+({\bf 15},{\bf 2})_{+1} +({\bf 35+1},{\bf 1})_{0}+({\bf 1},{\bf 3})_{0}+({\bf 15}',{\bf 2})_{-1}+({\bf 15},{\bf 1})_{-2}+({\bf 1},{\bf 2})_{-3}\,,
\end{equation}
with the ${\rm E}_{7(7)}$ generators splitting accordingly into
\begin{equation}
 \left\{ t^\alpha\,,\,\,\,t^{mnpq}\,,\,\,\,t^{\alpha\,mn}\,,\,\,\,t^m{}_n\,,\,\,\,t^\alpha{}_\beta\,,\,\,\,t_{\alpha\,mn}\,,\,\,\,
t_{mnpq}\,,\,\,\,t_\alpha \right\}\,.
\end{equation}
Next we write $\mathcal{M}_{MN}(x,y)$ in (\ref{gMExFT}) as $$\mathcal{M}(x,y)=\mathcal{V}_{{\rm IIB}}(x,y)\cdot \mathcal{V}_{{\rm IIB}}(x,y)^t\,,$$
where $\mathcal{V}_{{\rm IIB}}(x,y)$ is a representative of the coset ${\rm E}_{7(7)}/{\rm SU}(8)$ in the solvable gauge which is appropriate to the Type IIB theory \cite{Andrianopoli:1996zg,Cremmer:1997ct}:
\begin{equation}
\mathcal{V}_{{\rm IIB}}(x,y)=e^{t^\alpha\,B_\alpha}\cdot e^{\frac{1}{24}\,t^{mnpq}\,C_{mnpq}}
\cdot e^{\frac{1}{2}\,t^{\alpha\,mn}\,B_{\alpha\,mn}}\cdot \mathcal{V}_2\cdot \mathcal{V}_6\,,
\end{equation}
where $B_\alpha$ are the scalars dual in $D=4$ to $B_{\alpha\,\mu\nu}$, $C_{mnpq}$ are the internal components of the 4-form, $B_{\alpha\,mn}$ are the internal components of the 2-forms, $\mathcal{V}_6$ is the representative of ${\rm GL}(6,\mathbb{R})/{\rm SO}(6)$ and $\mathcal{V}_2$ that of ${\rm SL}(2,\mathbb{R})_{{\rm IIB}}/{\rm SO}(2)$, depending on the $D=10$ axion $C_0$ and dilaton $\phi$ fields. In our notations the doublet of ten dimensional 2-forms $B^\alpha_{(2)}$ is defined in terms of the NS-NS and R-R fields $B_{(2)},\,C_{(2)}$ as follows: $B^\alpha_{(2)}=\epsilon^{\alpha\beta}\,B_{\beta\,(2)}=(B_{(2)},\,C_{(2)})$.\footnote{In our conventions $\epsilon_{12}=\epsilon^{12}=+1$}\par
After having computed the matrix $\mathcal{M}(x,y)$ on the $\mathcal{N}=2$ vacua, the internal metric $G_{mn}(y)$, the internal components of the 2-forms ${B}_{mn}^\alpha=\epsilon^{\alpha\beta}\,B_{\beta\,mn}$, and the internal components of the 4-form $C_{mnpq}$ in the $D=10$ solution, can be computed as follows  \cite{Inverso:2016eet,Guarino:2019oct,Guarino:2020gfe}:
\begin{eqnarray}
  G^{mn}&=&G^{\frac{1}{2}}\mathcal{M}^{mn}  \nonumber \\
 {B}_{mn}^\alpha&=&G^{\frac{1}{2}}G_{mp}\epsilon^{\alpha \beta} \mathcal{M}^p{}_{ \beta n} \nonumber \\
 C_{mnpq} -\frac{3}{2}\epsilon_{\alpha \beta} {B}_{m\left[n\right.}^\alpha {B}_{\left. pq \right]}^\beta &=& -G^{\frac{1}{2}}G_{mr}\mathcal{M}^r{}_{npq} \nonumber \\
  m_{\alpha \beta} &=& \frac{1}{6} G \left( \mathcal{M}^{mn} \mathcal{M}_{\alpha m \,\beta m} + \mathcal{M}^m{}_{\alpha n } \mathcal{M}^n_{\beta m } \right)\,,
\end{eqnarray}
where $G\equiv {\rm det}(G_{mn})$. The matrix $m_{\alpha\beta}$ is an element of ${\rm SL}(2,\mathbb{R})_{{\rm IIB}}/{\rm SO}(2)$ and is defined as:
\begin{equation}
m_{\alpha\beta}\equiv (\mathcal{V}_2\cdot\mathcal{V}_2^t)_{\alpha\beta}=\frac{1}{{\rm Im}(\tau)}\left(\begin{matrix}|\tau|^2 & -{\rm Re}(\tau)\cr -{\rm Re}(\tau) & 1\end{matrix}\right)\,,\label{mab}
\end{equation}
where $\tau\equiv C_0+{i}\,e^{-\phi}$. In next section we shall perform the Kaluza-Klein analysis on the $\mathcal{N}=2$ vacua and in section \ref{uplift}, using the above formulas, we shall give the corresponding class of one-parameter $D=10$ solutions.

\section{The ${\cal N}=2$ Kaluza-Klein Spectrum from ExFT}
\label{KKExFT}

\subsection{ExFT spectroscopy}

The ExFT formulation of supergravity not only provides a powerful tool for uplifting lower-dimensional solutions, but
also for computing the Kaluza-Klein spectra around the resulting higher-dimensional backgrounds. The formalism has been set up in
\cite{Malek:2019eaz,Malek:2020yue} and here we briefly review the relevant formulas.
As a general structure, the Kaluza-Klein fluctuations around such a background are expressed as a product of the modes of the consistent truncation (\ref{gMExFT}) captured by the $U$ matrix, with a complete basis of functions on the compactification manifold.
In the case at hand, the basis of functions $\{{\cal Y}^\Sigma\}$ can be chosen to be a tensor product of
the scalar harmonics on the round $S^5$ with a standard Fourier expansion on $S^1$. More precisely,
we can use the following basis for harmonics
\begin{equation}
	{\cal Y}^\Sigma = \left\{ {\cal Y}^\sigma \otimes {\cal Y}^{(n)} \right\} \,,
	\label{harmYY}
\end{equation}
where
\begin{equation}
	{\cal Y}^\sigma = \left\{ {\cal Y}^{a},\, {\cal Y}^{a_1a_2} ,\, \ldots,\, {\cal Y}^{a_1 \ldots a_n},\, \ldots \right\} \,,
	\qquad
	a_i=1, \dots, 6\,,
\end{equation}
are the sphere harmonics on $S^5$ constructed as traceless symmetric products
${\cal Y}^{a_1 \ldots a_n} = {\cal Y}^{(\!(a_1} \ldots {\cal Y}^{a_n)\!)}$,
in terms of the fundamental harmonics, ${\cal Y}^a$, on $S^5$, which satisfy ${\cal Y}^a {\cal Y}^a = 1$,
and
\begin{equation}
	 {\cal Y}^{(n)} = \exp\left(\frac{2\pi\,i\,n}{T} \eta \right) \,,
\end{equation}
are the $S^1$ harmonics with periodicity $\eta = \eta + T$ of the $S^1$ coordinate $\eta$\,. The harmonics are related to the twist matrices from (\ref{gMExFT}) by a linear action
of generalized diffeomorphisms
\begin{equation}
	{\cal L}_{{{\cal U}_{M}}{}_{\vphantom{T}}} {\cal Y}^\Sigma
	= - {\cal T}_{M}{}^{\Sigma}{}_{\Omega}\, {\cal Y}^\Omega \,,
\label{TUT}
\end{equation}
with gauge parameters \eqref{calU}, see \cite{Malek:2019eaz,Malek:2020yue} for details. For the following, we simply note that this relation defines a set of constant matrices $({\cal T}_{M}){}^{\Sigma}{}_{\Omega}$, satisfying the algebra
\begin{equation}
	\left[ {\cal T}_{M},\, {\cal T}_{N} \right] = X_{MN}{}^{P} \,{\cal T}_{P} \,,
\end{equation}
which realizes the embedding tensor $X_{MN}{}^{P} $ as structure constants.
For the specific twist matrix $U_M{}^N(Y)$ defined above, the matrices $({\cal T}_{M}){}^{\Sigma}{}_{\Omega}$ acting on the harmonics \eqref{harmYY} have the following non-zero entries
\begin{equation}
	{\cal T}_{AB}{}^{c_1\ldots c_n}{}_{d_1 \ldots, d_n} =2\, n\,
	t_{[A}{}^{(\!(c_1} t^{\phantom{c}}_{B](\!(d_1} \delta_{d_2}^{c_2} \ldots \delta_{d_n)\!)}^{c_n)\!)} \,,
	\qquad {{\cal T}}^{67,(n)}{}_{(m)} = \frac{2\pi\,i\,n}{T} \,\delta^{(n)}_{(m)} \,,
	\label{TTT}
\end{equation}
where the matrix
\begin{equation}
t_A{}^c = \left\{
\begin{array}{cc}
\delta_{A,c} & A\le 5
\\
\delta_{A-2,c} & A=8 \\
0 & A = 6, 7
\end{array}
\right.\,,
\end{equation}
takes care of the embedding of the harmonics into the basis used to define $U_M{}^N(Y)$.

The fluctuation Ansatz of the  ExFT fields \eqref{fields} around an AdS$_{4}$ vacuum extends the Ansatz for the consistent truncation (\ref{gMExFT}) and
is given by \cite{Malek:2019eaz,Malek:2020yue}
\begin{equation} \label{eq:FluctAnsatz}
	\begin{split}
		g_{\mu\nu}(x,y) & = \rho^{-2} \, \Big( \mathring{g}_{\mu\nu}(x) + \displaystyle\sum_{\Sigma} \mathcal{Y}^{\Sigma} \, h_{\mu\nu , \Sigma}(x) \Big) \,, \\
		\mathcal{A}_{\mu}{}^{M}(x,y) & =  \rho^{-1} \, (U^{-1})_{\uA}{}^{M} \, \displaystyle\sum_{\Sigma} \mathcal{Y}^{\Sigma} \, A_{\mu}{}^{\uA , \Sigma}(x) \,, \\
		\mathcal{M}_{MN}(x,y) & = U_{M}{}^{\uA}
		U_{N}{}^{\uB} \, \Big( \delta_{\uA\uB} + {\cal P}_{\uA\uB,I}\displaystyle\sum_{\Sigma} \mathcal{Y}^{\Sigma}  j_{I , \Sigma}(x) \Big) \,,
	\end{split}
\end{equation}
where the Kaluza-Klein fluctuations for the metric, vector fields and scalars are labeled by $h_{\mu\nu,\Sigma}(x)$, $A_{\mu}{}^{\uA,\Sigma}$, and $j_{I,\Sigma} \in \mathfrak{e}_{7(7)} \ominus \mathfrak{su}(8)$, respectively. The twist matrix $U_{M}{}^{\uA}$ appearing in \eqref{eq:FluctAnsatz}
is obtained from the twist matrix from (\ref{gMExFT}) upon dressing with the scalar matrix of the four-dimensional supergravity, ${\cal V}_{{M}}{}^{\underline{A}} \in \textrm{E}_{7(7)}/\mathrm{SU}(8)$, evaluated at the scalar configuration specifying the ${\cal N}=2$ vacuum as
\begin{equation}
\label{dressed_U}
	U_{M}{}^{\uA}(y)=U_{M}{}^{N}(y) \, \mathcal{V}_{N}{}^{\underline{A}} \,.
\end{equation}
The scalar fluctuations in (\ref{eq:FluctAnsatz}) moreover appear under projection ${\cal P}_{\uA\uB,I}$, with $I=1,\ldots,70$, over the non-compact E$_{7(7)}$-generators resulting from the expansion of the group element $\mathcal{M}_{MN}$
on the 70-dimensional coset space E$_{7(7)}/{\rm SU}(8)$. The normalization of ${\cal P}_{\uA\uB,I}$ is not relevant since it drops out of the mass matrix when normalized relative to the scalar kinetic term.

Evaluating the ExFT field equations from \cite{Hohm:2013uia} with the fluctuation Ansatz \eqref{eq:FluctAnsatz} induces the mass matrices for the bosonic Kaluza-Klein spectrum which are expressed in terms of the embedding tensor
$X_{{MN}}{}^{{P}}$ from (\ref{XMNP}), and the matrices ${\cal T}$ from (\ref{TUT}), (\ref{TTT}),
both dressed by the scalar vielbein $\mathcal{V}_{{M}}{}^{\underline{A}}$ as
\begin{equation}
	\label{T-tensor}
	\begin{split}
	X_{\underline{AB}}{}^{\underline{C}} &= (\mathcal{V}^{-1})_{\uA}{}^{{M}} \, (\mathcal{V}^{-1})_{\uB}{}^{{N}}  \, X_{{MN}}{}^{{P}} \,  \mathcal{V}_{{P}}{}^{\underline{C}} \,, \\
	{\cal T}_{\uA}{}^{\Sigma}{}_\Omega &= (\mathcal{V}^{-1})_{\uA}{}^{{M}} \, {\cal T}_{M}{}^{\Sigma}{}_{\Omega} \,.
	\end{split}
\end{equation}
The mass matrices are obtained by linearizing the ExFT field equations with the fluctuation ansatz \eqref{eq:FluctAnsatz} \cite{Malek:2019eaz,Malek:2020yue}, and we give them in compact form as
\begin{equation}
\begin{split}
\mathbb{M}^{\operatorname{(spin-2)}}_{\Sigma\Omega} &= -({\cal T}_{\underline{A}}{\cal T}_{\underline{A}})_{\Sigma\Omega}
\,,
\\[1ex]
\mathbb{M}^{{\rm (vector)}}_{\underline{A}\Sigma,\underline{B}\Omega}
&= (\Pi \Pi^T)_{\underline{A}\Sigma,\underline{B}\Omega}
\,,
\\[1ex]
\mathbb{M}^{{\rm (scalar)}}_{I\Sigma,J\Omega}
&=
{\cal M}^{(0)}_{IJ}\,\delta_{\Sigma\Omega}
+{\delta}_{IJ}\,\mathbb{M}^{\operatorname{(spin-2)}}_{\Sigma\Omega}
+{\cal N}_{IJ}{}^{\underline{C}}\,{\cal T}_{\underline{C},\Sigma\Omega}
-\tfrac16(\Pi^T \Pi)_{I\Sigma,J\Omega}
\,.
\end{split}
\label{KKmasses}
\end{equation}
The tensors appearing in these expressions are given by
\begin{equation}
\begin{split}
{\Pi}_{\underline{A}\Sigma,I\Omega}
=~&
\delta_{\Sigma\Omega}\,
X_{\underline{AC}}{}^{\underline{D}} \, {\cal P}_{\underline{CD},I}
- 12\,{\cal P}_{\underline{AD},I}\,  {\cal T}_{\underline{D}\,\Omega\Sigma}
\,,
\\[1ex]
{\cal N}_{IJ}{}^{\underline{C}}
=~&
-4\left(X_{\underline{CA}}{}^{\underline{B}}+12\,X_{\underline{AB}}{}^{\underline{C}}\right)
{\cal P}_{\uA\uD}{}^{[I}
{\cal P}_{\uB\uD}{}^{J]}
\,,
\\[1ex]
{\cal M}^{(0)}_{IJ} =~&
  \tfrac17
 \left(
 7\,X_{\uA\uE}{}^{\uF} X_{\uB\uF}{}^{\uE}+
 X_{\uA\uE}{}^{\uF} X_{\uB\uE}{}^{\uF} + X_{\uE\uA}{}^{\uF} X_{\uE\uB}{}^{\uF} + X_{\uE\uF}{}^{\uA} X_{\uE\uF}{}^{\uB}  \right)
 {\cal P}_{\uA\uD,I}\, {\cal P}_{\uB\uD,J}
  \\&{}
+  \tfrac27
  \left( X_{\uA\uC}{}^{\uE} X_{\uB\uD}{}^{\uE} - X_{\uA\uE}{}^{\uC} X_{\uB\uE}{}^{\uD} - X_{\uE\uA}{}^{\uC} X_{\uE\uB}{}^{\uD} \right)
 {\cal P}_{\uA\uB,I}\, {\cal P}_{\uC\uD,J}
\,.
\end{split}
\end{equation}
In particular, the matrix ${\cal M}^{(0)}_{IJ}$ is the mass matrix
derived from the scalar potential of $D=4$ supergravity,
describing the masses of the 70 scalars at the lowest Kaluza-Klein level.
The corresponding mass formulas for the fermionic sector have
been worked out in \cite{Cesaro:2020soq}.

\subsection{The Kaluza-Klein spectrum around the ${\cal N}=2$ backgrounds} \label{s:Spectrum}

Before applying the ExFT technology to the family of ${\cal N}=2$ vacua of interest, let us first work out
to which extent the structure of the spectrum is constrained from the representation structure
of the underlying supergroup ${\rm OSp}(\mathcal{N}|4)$. The generic supermultiplet of this group is
of long type
\be
{L\bar{L}[J]^{(R)}_\Delta}\,,\qquad
J=0, \tfrac12, 1\,,
\label{longLL}
\ee
with $J$ referring to the Lorentz spin of the highest weight state (HWS), such that its different values in (\ref{longLL})
correspond to the long vector, gravitino, and graviton multiplets, respectively. Labels $\Delta$, and $R$ refer
to the conformal dimensions and the ${\rm U}(1)_R$ R-symmetry charge of the HWS, respectively.
Unitarity implies a lower bound for the conformal dimension
\be
\Delta ~\ge~ 1+ |R| + J
\,.
\label{unitarity}
\ee
When the bound is saturated, the long multiplet
decomposes into shortened multiplets. We refer to appendix~\ref{supermult} for notation and details on these multiplets
and the shortening patterns.

The presence of an $S^1$ factor in our backgrounds implies that all masses continuously depend on the inverse circle radius.
Only at generic values of the radius, the spectrum thus necessarily assembles into long multiplets (\ref{longLL}). At specific values
of the inverse radius (and in particular for the zero modes on the circle) some of the long multiplets fall to the unitarity bound (\ref{unitarity})
and decompose into shortened multiplets.

To make the results explicit, let us recall the character/partition function of the long multiplets (\ref{longLL}), given by
\begin{equation}\begin{split}
Z_{L\bar{L}[0]_\Delta^{(R)}} &= Z_{0}[\Delta,R] \equiv
t^\Delta u^R\,\big(1-\sqrt{t}\,  \tfrac{\sqrt{z}}{u}\big)
\big(1-\sqrt{t}\,  \tfrac{1}{\sqrt{z}\,u}\big)
\big(1-\sqrt{t}\,  \sqrt{z}\,u\big)
\big(1-\sqrt{t}\,  \tfrac{u}{\sqrt{z}}\big)
\,,
\\
Z_{L\bar{L}[\frac12]_\Delta^{(R)}} &= Z_{\frac12}[\Delta,R] \equiv
-t^\Delta u^R\,\big(\sqrt{z}+\tfrac1{\sqrt{z}}\big)\,
\big(1-\sqrt{t}\,  \tfrac{\sqrt{z}}{u}\big)
\big(1-\sqrt{t}\,  \tfrac{1}{\sqrt{z}\,u}\big)
\big(1-\sqrt{t}\,  \sqrt{z}\,u\big)
\big(1-\sqrt{t}\,  \tfrac{u}{\sqrt{z}}\big)
\,,
\\
Z_{L\bar{L}[1]_\Delta^{(R)}} &= Z_{1}[\Delta,R] \equiv
t^\Delta u^R\,\big(z+1+\tfrac1{z}\big)\,
\big(1-\sqrt{t}\,  \tfrac{\sqrt{z}}{u}\big)
\big(1-\sqrt{t}\,  \tfrac{1}{\sqrt{z}\,u}\big)
\big(1-\sqrt{t}\,  \sqrt{z}\,u\big)
\big(1-\sqrt{t}\,  \tfrac{u}{\sqrt{z}}\big)
\,.
\label{charactersLong}
\end{split}\end{equation}
c.f.  appendix  \ref{supermult}.
Here, exponents of $t$, $u$, and $z$ count the conformal dimension, R-charge, and Lorentz spin, respectively.
Following the above discussion, the partition function
for the full Kaluza-Klein spectrum can thus be written in the form
\begin{equation}
Z_{\rm KK} =
\nu_0\,Z_{0}[0,0] +
\nu_{1\!/\!2}\,Z_{\frac12}[0,0] +
\nu_1\,Z_{1}[0,0]
\;,
\label{LKK}
\end{equation}
with the characters $\nu_0$, $\nu_{1\!/\!2}$, $\nu_1$,
carrying the HWS of the long multiplets.

Except for the masses, the remaining quantum numbers of the spectrum can be inferred from the fluctuation ansatz (\ref{eq:FluctAnsatz}),
upon multiplying the fields of ${\cal N}=8$ supergravity with the tower of scalar harmonics. To this end, let us note that
the ${\rm U}(2)$ symmetry, preserved at the $\chi=0$ vacuum, is embedded into the ${\rm SO}(6)$ part of the gauge group,
such that the gravitini decompose as
\begin{equation}\begin{split}
{\bf 8}_s \longrightarrow~&
2\times [0]_{+1}+2\times [0]_{-1}+2\times [\tfrac12]_0
\,,\\[1ex]
\mbox{i.e.}\quad
\nu_8 =~& 2\,u+\tfrac{2}{u} + 2\,\sqrt{x}+\tfrac{2}{\sqrt{x}}\,,
\end{split}\end{equation}
where $x$ counts the ${\rm U}(1)\subset {\rm SU}(2)$ charges.
From this, the ${\rm U}(2)$ representation content of the full ${\cal N}=8$ supergravity multiplet can be deduced as
\begin{equation}\begin{split}
\mbox{graviton :}\quad {\bf 28} :&~~  {\bf 1}
\,,\\
\mbox{gravitini :}\quad {\bf 28} :&~~  {\bf 8}_s
\,,\\
\mbox{vectors :}\quad {\bf 28} :&~~  {\bf 8}_s  \wedge {\bf 8}_s
\,,\\
\mbox{spin-$\frac12$ fermions :}\quad {\bf 56} :&~~ {\bf 8}_s  \wedge{\bf 8}_s  \wedge{\bf 8}_s
\,,\\
\mbox{scalars :}\quad {\bf 70} :&~~ {\bf 8}_s  \wedge{\bf 8}_s  \wedge{\bf 8}_s  \wedge{\bf 8}_s
\,.
\label{spectrumN8}
\end{split}\end{equation}
The $S^5$ sphere harmonics in turn decompose as
\begin{equation}\begin{split}
{\bf 6} \longrightarrow~&
2\times [0]_{0}+ [\tfrac12]_{+1} + [\tfrac12]_{-1}
\,,\\[1ex]
\mbox{i.e.}\quad
\nu_6 =~& 2+ \sqrt{x}\,u+\tfrac{u}{\sqrt{x}}
+\tfrac{1}{\sqrt{x} u}+\tfrac{\sqrt{x}}{ u}\,,
\label{harm6}
\end{split}\end{equation}
under ${\rm U}(2)$. The full Kaluza-Klein spectrum then is obtained by multiplying
(\ref{spectrumN8}) with the symmetric tower of $S^5$ harmonics (\ref{harm6}) and the tower of $S^1$ harmonics, the latter amounting to a standard Fourier expansion. Comparing the result to the general form (\ref{LKK}), we may read off the characters $\nu_J$ except for the conformal dimensions, i.e.\ setting $t=1$, and find
\begin{equation}\begin{split}
\nu_0\big|_{t=1} &=
\frac{(1-q^2)\,\big(x+3+\tfrac1{x}\big)}{(1-q)^2\,
\big(1-q\,  \tfrac{\sqrt{x}}{u}\big)
\big(1-q\,  \tfrac{1}{\sqrt{x}\,u}\big)
\big(1-q\,  \sqrt{x}\,u\big)
\big(1-q\,  \tfrac{u}{\sqrt{x}}\big)}
\,\frac{1+s}{1-s}
\,,\\
\nu_{1\!/\!2}\big|_{t=1} &=
\frac{2\,(1-q^2)\,\big(\sqrt{x}+\tfrac{1}{\sqrt{x}}\big)}
{(1-q)^2\,
\big(1-q\,  \tfrac{\sqrt{x}}{u}\big)
\big(1-q\,  \tfrac{1}{\sqrt{x}\,u}\big)
\big(1-q\,  \sqrt{x}\,u\big)
\big(1-q\,  \tfrac{u}{\sqrt{x}}\big)}
\,\frac{1+s}{1-s}
\,,\\
\nu_1\big|_{t=1} &=
\frac{1-q^2}{(1-q)^2\,
\big(1-q\,  \tfrac{\sqrt{x}}{u}\big)
\big(1-q\,  \tfrac{1}{\sqrt{x}\,u}\big)
\big(1-q\,  \sqrt{x}\,u\big)
\big(1-q\,  \tfrac{u}{\sqrt{x}}\big)}
\,\frac{1+s}{1-s}
\,.
\label{chis}
\end{split}\end{equation}
Here, exponents of $q$, $s$, count levels for the $S^5$ and the $S^1$ harmonics, respectively.
The $S^1$ factor $\frac{1+s}{1-s}$ simply encodes the fact that at $S^1$--level $n>0$
the harmonics (Fourier modes) are complex.

Representation theory alone thus determines the Kaluza-Klein spectrum to be of the form
 (\ref{LKK}), (\ref{chis}). The last and central information which completes this spectrum is the assignment of conformal dimensions/masses to all the states. It is at this step, that the ExFT technology described in the previous subsection becomes relevant.
After evaluating the mass matrices (\ref{KKmasses})
for the spin-2, the vector and the scalar fields, respectively,
we can extract a general formula for the conformal dimensions $\Delta$ of the
HWS of the supermultiplets, counted by \eqref{chis}, as
\begin{equation}\begin{split}
\Delta =~&  \tfrac12+
\sqrt{\tfrac{17}{4}+ \tfrac12 R^2
-J(J+1)   - 2 k (k + 1) +\ell(\ell + 4) +4 \left(\tfrac{\pi n}{T} -  j \chi\right)^2}
\,
\\[1ex]
&
\mbox{for a HWS of type\;\;} q^\ell\,s^n\,u^R\,x^j\,z^J\;\;
\mbox{and SU(2) spin}\; k
\;.
\label{Delta}
\end{split}\end{equation}
The conformal dimensions inside the multiplets then follow from the multiplet structure
(\ref{charactersLong}).
Combining (\ref{LKK}), (\ref{chis}), and (\ref{Delta}) thus produces the full Kaluza-Klein spectrum.
We stress again, that for some multiplets, the conformal dimensions determined from (\ref{Delta})
may saturate the unitarity bound (\ref{unitarity}), such that the corresponding long multiplets appearing in the expansion (\ref{LKK})
split into shortened multiplets.

\begin{figure}[tb]
   \centering
   \includegraphics[width=16.5cm]{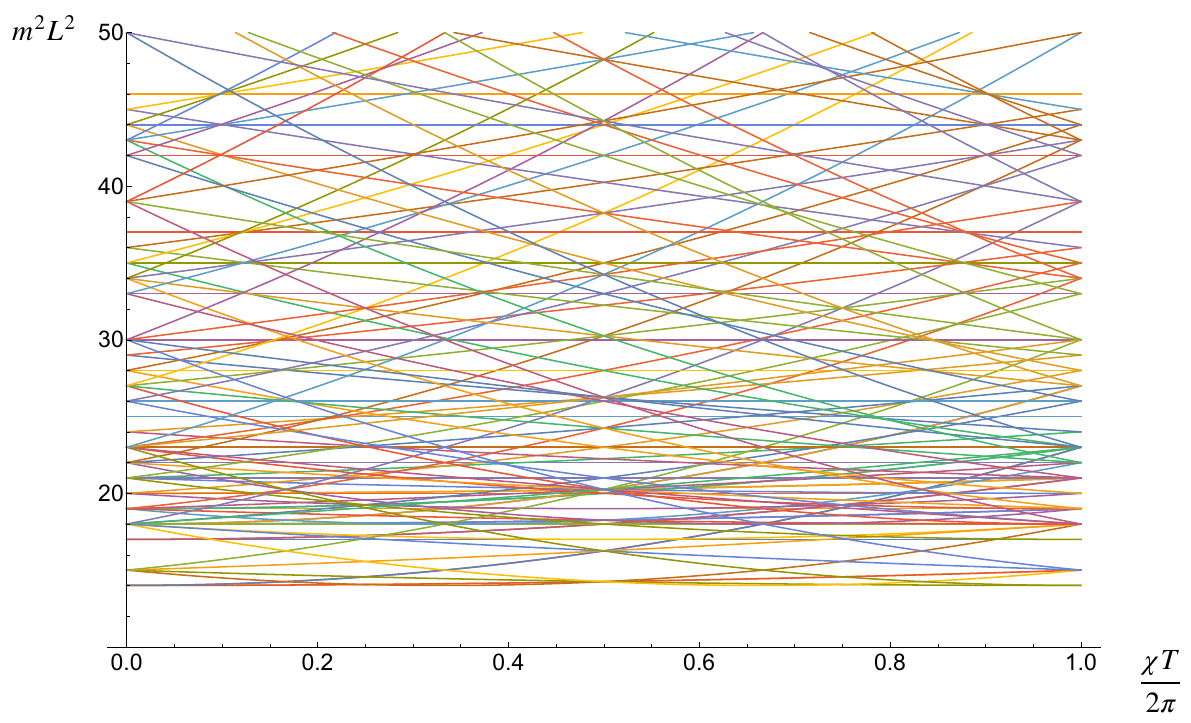}
   \caption{Spin-2 masses at level $\ell =3$, as a function of $\chi$. $L$ denotes the AdS$_4$ radius.}
   \label{fig:lev3}
\end{figure}

The mass formula (\ref{Delta}) explicitly shows that a non-vanishing $\chi\not=0$ breaks ${\rm SU}(2)$ by terms
proportional to the ${\rm U}(1)\subset {\rm SU}(2)$ charge. Moreover, it exhibits and interesting
interplay between the $\chi$-dependence and the $S^1$--level $n$: all masses receive correction terms proportional
to
\be
\left(\tfrac{\pi n}{T} -  j \chi\right)^2
\;.
\label{mmnj}
\ee
In particular, this allows to deduce that the full mass
spectrum is mapped onto itself under shifts $\chi\rightarrow \chi+\frac{2\pi}{T}$\,.
Indeed, upon switching on $\chi$, the ${\rm SU}(2)$ representations at a given $S^1$--level $n$ break up into their ${\rm U}(1)$
constituents which then at $\chi=\frac{2\pi}{T}$ recombine (over various levels)
into a copy of the original ${\rm SU}(2)$ representations.
More precisely, a state of ${\rm SU}(2)$ spin $k$ at level $n$ and generic value of the deformation parameter $\chi$ breaks up into
the $2k+1$ states of ${\rm U}(1)$ charge
\be
j\in
\{-k, -k+1, \dots, k\}
\,,
\ee
with conformal dimensions $\Delta_\chi$ given by (\ref{Delta}), thus
deformed by contributions in (\ref{mmnj}). For $\chi=\frac{2\pi}{T}$ on the other hand, every level
$\tilde{n}$ in the range
\be
\tilde{n} \in \{|n-k|, |n-k+1|, \dots, n+k\}
\,,
\ee
carries a state of conformal dimension $\Delta_0$ which recombine into a spin $k$ representation of a (newly enhanced)
${\rm SU}(2)$ symmetry.
As an illustration, Figure~\ref{fig:lev3} depicts the spectrum of spin-2 masses at fixed $S^5$--level ${\ell}=3$.
It shows the breaking and recombining of the spin-2 states as a function of the deformation parameter $\chi$ running from
$0$ to $\frac{2\pi}{T}$. The spectra at the two endpoints $\chi=0$ and $\chi=\frac{2\pi}{T}$ are identical.

It is also instructive to illustrate this pattern at the lowest $S^5$--level $\ell=0$.
At this level, the spectrum combines into supermultiplets
\begin{equation}\begin{split}
&\!\!\!\!\!\!\!\!
4\times L\bar{L}[0]^{(0)}_{\frac12+\sqrt{\frac{17}4+\frac{4\pi^2n^2}{T^2}}}
\oplus \;\; 2\times  L\bar{L}[0]^{(0)}_{\frac12+\sqrt{\frac{1}4+\left(\frac{2\pi n}{T}\pm2\chi\right)^2}}
\oplus
\;\;
2\times  L\bar{L}[0]^{(0)}_{\frac12+\sqrt{\frac{1}4+\frac{4\pi^2 n^2}{T^2}}}
\\[1ex]
&
\oplus \;\;4\times  L\bar{L}[\tfrac12]^{(0)}_{\frac12+\sqrt{2+\left(\frac{2\pi n}{T}\pm\chi\right)^2}}
\oplus \;\;2\times  L\bar{L}[1]^{(0)}_{\frac12+\sqrt{\frac94+\frac{4\pi^2 n^2}{T^2}}}
\;,
\label{L0n}
\end{split}\end{equation}
for $S^1$--level $n>0$,
accompanied by (\ref{level00}) at level $n=0$\,. At $\chi=0$, some of the conformal dimensions in (\ref{L0n}) degenerate with the corresponding
supermultiplets joining into irreducible ${\rm SU}(2)$ representations  of spin ${\bf [1]}$ and ${\bf [\tfrac12]}$, respectively.
At level 0, this moreover induces the multiplet shortening (\ref{shorteningLL10})
with the two arising massless vector multiplets $A_2\bar{A}_2[0]^{(0)}_1$
manifesting the symmetry enhancement ${\rm U}(1)\rightarrow{\rm SU}(2)$,
as discussed in the previous section.
In contrast, at $\chi=\frac{2\pi}{T}$, those additional massless vector multiplets
arise from multiplet shortening of the second multiplet in (\ref{L0n}) at level $n=2$.
This is an explicit realisation of a (bosonic version of the) space invader scenario encountered in
other compactifications \cite{Duff:1986hr}, in which massive fields from higher Kaluza-Klein
levels turn into massless gauge fields.

However, the structure of the spectrum (\ref{L0n}) shows an even more remarkable structure at the intermediate
value $\chi=\frac{\pi}{T}$\,. At this value, multiplet shortening of the second multiplet in (\ref{L0n}) now at level $n=1$
gives rise to two additional massless vector multiplets which reveal another ${\rm SU}(2)$ symmetry
enhancement at this point. In contrast with the symmetry enhancement at $\chi=\frac{2\pi}{T}$, the full
Kaluza-Klein spectrum at this intermediate point is different from the one at $\chi=0$. A closer look at the
$\chi$-dependence (\ref{mmnj}) of the masses shows that under $\chi\rightarrow\chi+\frac{\pi}{T}$,
the spectrum of states of integer ${\rm SU}(2)$ spin maps into itself whereas the states of half-integer
${\rm SU}(2)$ spin acquire different masses. This is also visible in Figure~\ref{fig:lev3} with
the degeneracies due to the symmetry enhancement to an inequivalent spectrum at the intermediate point $\chi=\frac{\pi}{T}$\,.
It is worth pointing out that the truncation of the level 0 spectrum (\ref{level00}) to integer ${\rm SU}(2)$ spin amounts
to truncating the four-dimensional ${\cal N}=8$ supergravity to a half-maximal ${\cal N}=4$ theory.

In section~\ref{uplift}, we will discuss the higher-dimensional origin responsible for these patterns.
\subsubsection{Symmetries of the Kaluza-Klein spectrum}
Inspection of the Kaluza-Klein spectrum shows the following two symmetries:
\begin{align}
&\chi \rightarrow \chi+\frac{2\pi}{T}\,\,\,\,,\,\,\,\,\,\,n\rightarrow n+2j\,,\label{sym1}\\
&\chi \rightarrow -\chi\,\,\,\,,\,\,\,\,\,\,j\rightarrow -j\,.\label{sym2}
\end{align}
The above symmetries combine into a reflection symmetry of the spectrum in the $\chi=\pi/T$ vertical line:
 \begin{align}
&\chi \rightarrow \frac{2\pi}{T}-\chi\,\,\,\,,\,\,\,\,\,\,n\rightarrow n-2j\,\,\,\,,\,\,\,\,\,\,j\rightarrow -j\,,
\end{align}
 which is manifest in Figure~\ref{fig:lev3}.\par
 Later in Section \ref{GOTIS}, we will give a characterization of the symmetries (\ref{sym1}) and (\ref{sym2}) in terms of the geometric properties of an elliptic fibration within the internal manifold. In this construction $\chi$ will be identified with the real part of the complex structure modulus of a torus fibered over $S^2$. The symmetry (\ref{sym1}) will then be interpreted as the Dehn twist, see Subsection \ref{s:CStructure} on the fiber, which can be reabsorbed in a globally well defined reparametrization of the deformed $S^3$, while (\ref{sym2}) as the effect of a parity transformation on the same fiber, see Subsection \ref{finalchi}.

\subsection{Multiplet shortening}

As discussed above, at  $\chi=0$, the symmetry enhances according to
${\rm U}(1)^2 \rightarrow {\rm U}(2)$. At the same time, at these values,
the conformal dimensions (\ref{Delta}) of several supermultiplets hit the
unitarity bound (\ref{unitarity}) and the generic long multiplets split up
into shortened multiplets according to the patterns reviewed in appendix~\ref{supermult}.
Explicitly, combining the saturation of the unitarity bound
\be
\Delta = 1+ |R| + J
\,,
\label{bound}
\ee
with the formula (\ref{Delta})
translates into the condition
\begin{equation}
8    +2 \ell(\ell + 4)
=
(|r|+2J)(|r|+2J+2)
+4 k (k + 1)
\;.
\end{equation}
Combining this with the bounds derived from the specific characters
(\ref{chis}), we conclude that multiplet shortening appears for the multiplets
whose HWS charges satisfy
\begin{equation}
|R| = \ell\,,\quad
k = 1+ \tfrac12\ell  - J
\;.
\label{inq}
\end{equation}
This reveals six series of long multiplets which sit on the unitarity bound and
each decompose into semi-short multiplets
according to (\ref{shortening})
\begin{equation}\begin{split}
[\tfrac{\ell}{2}] \otimes L\bar{L}[1]^{(\pm\ell)}_{\ell+2} &~~\longrightarrow~~
[\tfrac{\ell}{2}] \otimes
\left\{
\begin{array}{l}
L\bar{A}_1[1]^{(\ell)}_{\ell+2} + L\bar{A}_1[\tfrac12]^{(\ell+1)}_{\ell+5/2}
\\[2ex]
A_1\bar{L}[1]^{(-\ell)}_{\ell+2} + A_1\bar{L}[\tfrac12]^{(-\ell-1)}_{\ell+5/2}
\end{array}
\right.
\,,\\[4ex]
[\tfrac{\ell+1}{2}] \otimes L\bar{L}[\tfrac12]^{(\pm\ell)}_{\ell+\frac32} &~~\longrightarrow~~
[\tfrac{\ell+1}{2}] \otimes
\left\{
\begin{array}{l}
L\bar{A}_1[\tfrac12]^{(\ell)}_{\ell+\frac32} + L\bar{A}_2[0]^{(\ell+1)}_{\ell+2}
\\[2ex]
A_1\bar{L}[\tfrac12]^{(-\ell)}_{\ell+\frac32} + A_2\bar{L}[0]^{(-\ell-1)}_{\ell+2}
\end{array}
\right.
\,,\\[4ex]
[\tfrac{\ell+2}{2}] \otimes L\bar{L}[0]^{(\pm\ell)}_{\ell+1} &~~\longrightarrow~~
[\tfrac{\ell+2}{2}] \otimes
\left\{
\begin{array}{l}
L\bar{A}_2[0]^{(\ell)}_{\ell+1} + L\bar{B}_1[0]^{(\ell+2)}_{\ell+2}
\\[2ex]
A_2\bar{L}[0]^{(-\ell)}_{\ell+1} + B_1\bar{L}[0]^{(-\ell-2)}_{\ell+2}
\end{array}
\right.
\,,
\end{split}\end{equation}
at level $\ell>0$.

Similar multiplet shortening occurs at $\chi=\frac{\pi}{T}$\,.
More remarkably, multiplet shortening in fact happens at every value of $\chi$ that is a
rational multiple of $\frac{2\pi}{T}$\,.
More precisely, at
\begin{equation}
\chi= \frac{p}{q}\frac{2\pi}{T}\,,
\qquad
p, q \in \mathbb{N}
\,,
\end{equation}
shortening occurs for the multiplets whose HWS have U(1) charge
\begin{equation}
 j  = \frac{q \,n}{2\,p} \in \tfrac12 \mathbb{N}
\,.
\end{equation}
These multiplets appear at $S^1$ levels $n$ that are integer multiples of $p$, i.e.\ $n=m p$ with $m\in\mathbb{N}$.
We stress however, that the resulting shortened multiplets are not necessarily protected, as they can potentially
recombine again into the original long multiplets. It remains an open question to what extent they can
be recovered in the dual conformal field theory.

\section{The Type IIB Uplift of the 1-Parameter $\mathcal{N}=2$ Vacua}\label{uplift}
Just as in the $\chi=0$ case, the $D=10$ dimensional solution corresponding to the 1-parameter family of $\mathcal{N}=2$ vacua has the geometry of AdS$_4\times S^5\times S^1$, where $S^5$ denotes here a deformed five-sphere. As we shall see this family is locally related to the $\chi=0$ solution by a coordinate transformation involving the coordinates of $S^1$ and a squashed $S^3$ within $S^5$.

\subsection{Geometry of the Internal Space}\label{GOTIS}
We locally parametrize $S^5$ by coordinates $\theta,\,\varphi,\,\alpha,\,\beta,\,\gamma$ and $S^1$ by the coordinate $\eta$, with the following ranges
\begin{equation}
  \label{angcoord}
0\leq \eta < T\,,\,\,\,0 \leq \theta \le \frac{\pi}{2}\,,\,\,\,0 \leq \varphi < 2\pi\,,\,\,\,0 \leq \alpha \le 2\pi\,,\,\,\,0 \leq \beta \le \pi\,,\,\,\,0 \leq \gamma + \frac{\pi}{2} < 4\pi\,.
\end{equation}
The coordinates $\theta,\,\varphi$ parametrize an $S^2$ within $S^5$, while $\alpha,\,\beta,\,\gamma$ parametrize an $S^3$ within the same manifold.

We begin by describing the internal geometry for the $\chi = 0$ solution, before explaining how it is modified when $\chi \neq 0$. To understand the effect of $\chi \neq 0$, it is sufficient to focus on $S^3$ and $S^1$. In the full solution, discussed in section \ref{s:Metric}, the $S^3$ is fibred over $S^2$ in such a way that, for $\chi = 0$, only an $\SU(2) \times \U(1)' \subset \SU(2) \times \SU(2)'$ isometry of $S^3$ remains.

Hence it is convenient to describe $S^3$ using the isomorphism ${\rm SU}(2) \simeq S^3$, given by
\begin{equation}
	\label{groupel}
	g(\alpha,\beta,\gamma)\equiv \left(\begin{matrix}z_1 & z_2 \cr -\bar{z}_2 & \bar{z}_1\end{matrix}\right)=\left(
	\begin{array}{cc}
		e^{\frac{1}{2} i \, \left(\alpha +\gamma +\pi/2\right)} \cos \left(\frac{\beta }{2}\right)
		& e^{\frac{1}{2}  i \, \left(-\alpha +\gamma+\pi/2\right)} \sin \left(\frac{\beta
		}{2}\right) \\
		-e^{-\frac{1}{2}  i \, \left(-\alpha +\gamma+\pi/2
			\right)} \sin \left(\frac{\beta
		}{2}\right) & e^{-\frac{1}{2}  i \, \left(\alpha
			+\gamma+\pi/2 \right)} \cos
		\left(\frac{\beta }{2}\right) \\
	\end{array}
	\right)\,,
\end{equation}
where $z_1,\,z_2$ satisfy $|z_1|^2+|z_2|^2=1$ and define the embedding of $S^3$ in $\mathbb{C}^2$. The map from \eqref{angcoord} to the $y^m$ coordinates in \eqref{Ydec} is
\begin{equation}
	\frac{y^1}{\cos{\theta}}= {\rm Re}(z_1)\,, \qquad \frac{y^4}{\cos{\theta}}= {\rm Im}(z_1)\,, \qquad \frac{y^5}{\cos{\theta}}= {\rm Re}(z_2) \,,
\end{equation}
and
\begin{equation}
	y^2=\cos{\phi}\sin{\theta}\,, \qquad y^3=\sin{\phi}\sin{\theta}\,, \qquad \tilde{y}=\sinh{\eta} \,.
\end{equation}
For $\chi = 0$, we can express the squashed $S^3$ metric in terms of the left-invariant 1-forms $\sigma^i$, $i = 1, \ldots, 3$, defined as
\begin{equation}
	g^{-1}dg\,=\,\sum_{i=1}^3 \sigma^i ( i \,\boldsymbol{\sigma}^i)\,,
\end{equation}
where $\boldsymbol{\sigma}^i$ are the three Pauli matrices. The $\sigma^i$ satisfy the Maurer-Cartan equations
\begin{equation}
	d\sigma^i-\epsilon^{ijk}\,\sigma^j\wedge \sigma^k=0\,,\label{MC}
\end{equation}
with $\epsilon^{ijk} = \pm 1$ the structure constants of ${\rm SU}(2)$. Evaluating the Maurer-Cartan forms in terms of the coordinates \eqref{angcoord}, we find
\begin{equation}\begin{split}
\sigma^1&=\frac{1}{2} (\text{d$\gamma $} \cos (\alpha ) \sin
   (\beta )-\text{d$\beta $} \sin (\alpha ))\,,\qquad \sigma^2=\frac{1}{2} (\text{d$\beta $} \cos (\alpha
   )+\text{d$\gamma $} \sin (\alpha ) \sin (\beta
   ))\,,\\
   \sigma^3&=\frac{1}{2} (\text{d$\alpha $}+\text{d$\gamma $}
   \cos (\beta ))\,.
\end{split}\end{equation}

The dependence of the internal metric and the other fields, in the $\chi=0$ solution, on the point in $S^3$ is expressed in terms of $\sigma^i$ and thus the solution features an ${\rm SU}(2)$ symmetry group acting from the left on $g(\alpha,\beta,\gamma)$ and thus leaving $\sigma^i$ invariant.
Due to the squashing of the $S^3$ geometry, only a ${\rm U}(1)' $ subgroup of the ${\rm SU}(2)'$ group acting on $g(\alpha,\beta,\gamma)$ from the right is a symmetry of the $\chi=0$ solution. In fact the group ${\rm U}(1)' $ coincides with the $\mathcal{N}=2$ $R$-symmetry group, previously denoted by ${\rm U}(1)_R $.

For $\chi\neq 0$ the solution features a fibration of $S^3$ over $S^1$ in which a point of $S^1$ is associated with an $S^3$ parametrized by coordinates $(\alpha', \beta', \gamma')$ which define the following ${\rm SU}(2)$-element
\begin{equation}
g(\alpha', \beta', \gamma')=\hat{g}(\alpha,\beta,\gamma,\eta)\equiv h(\eta)\cdot g\left(\alpha,\beta,\gamma\right)\,,\label{hatg}
\end{equation}
where
\begin{equation}
	h(\eta)\equiv\left(
	\begin{array}{cc}
		\cos (\eta  \chi ) & \sin (\eta  \chi ) \\
		-\sin (\eta  \chi ) & \cos (\eta  \chi ) \\
	\end{array}
	\right) \in {\rm SU}(2) \,.\label{chitwist}
\end{equation}
The relation \eqref{hatg} defines the transition function on the $S^3$ fiber when changing chart on $S^1$ and introduces a monodromy on the same fiber as $\eta\rightarrow \eta+T$, represented by the left action of the element $h(T)=h(\eta)^{-1}h(\eta+T)$ in ${\rm SU}(2) $. Then, the total space of the 4-dimensional fiber-bundle, with fiber $S^3$ and base $S^1$, is given by the quotient space $S^3\times [0,\,T]/\sim$ where the identification $\sim$ is defined as follows
\begin{equation}
\left[g\left(\alpha,\beta,\gamma\right),\,\eta=0\right]\,\sim\, \left[h(T)\cdot g\left(\alpha,\beta,\gamma\right),\,\eta=T\right]\,.
\end{equation}

The presence of this monodromy further breaks the ${\rm SU}(2)$ isometry, that the squashed $S^3$ has for $\chi = 0$, to the subgroup of ${\rm SU}(2)$ commuting with $h(T)$. In general, this subgroup is given by ${\rm U}(1)$, while the isometry $\U(1)'$ coming from the right-action remains of course unbroken by $h(T)$. The values $\chi = \frac{2\pi}{kT}$ are particularly interesting since then the element $h(T)$ generates the cyclic group $\mathbb{Z}_k$. For $k=1$, the quotient is trivial so that four-dimensional manifold, like the $\chi = 0$ case, is a direct product of $S^3 \times S^1$ with isometry group ${\rm SU}(2)$. This will be further clarified in  section \ref{s:CStructure}, where we show that for $k = 1$, i.e.\ $\chi = \frac{2\pi}T$, the solution is equivalent to the $\chi = 0$ one.

For $k=2$, namely $\chi=\frac{\pi}{T}$, the twist commutes with all of ${\rm SU}(2) $ since the $\mathbb{Z}_2$ group it generates is the center of ${\rm SU}(2) $. Thus for $k=1,2$ the ${\rm U}(1) $ isometry is enhanced to ${\rm SU}(2) $. This explains the symmetry enhancement, for those special values of $\chi$, observed in section \ref{KKExFT} by inspection of the Kaluza-Klein spectrum.

Concerning the geometric description of the internal space, we observe that \emph{locally} $h(\eta)$ in \eqref{hatg} can be absorbed into a coordinate transformation:
\begin{equation} \label{eq:CoordChange}
\{\alpha,\beta,\gamma,\eta\}\,\rightarrow \,\,\,\{\alpha'(\alpha,\beta,\gamma,\eta),\,\beta'(\alpha,\beta,\gamma,\eta),\,\gamma'(\alpha,\beta,\gamma,\eta),\,\eta'=\eta\}\,,
\end{equation}
where $\alpha'(\alpha,\beta,\gamma,\eta),\,\beta'(\alpha,\beta,\gamma,\eta),\,\gamma'(\alpha,\beta,\gamma,\eta)$ are defined by the solution to the matrix equation
\begin{equation}
  \label{repara}
g(\alpha',\beta',\gamma')=\hat{g}(\alpha,\beta,\gamma,\eta)\,.
\end{equation}
Therefore, we can express the $\chi \neq 0$ solution by computing the new left-invariant 1-forms $\hat{\sigma}^i$ associated with $\hat{g}(\alpha,\beta,\gamma,\eta)$ or, equivalently, $g(\alpha',\beta',\gamma')$, as
\begin{equation}
	\hat{g}^{-1}d\hat{g}=\sum_{i=1}^3 \hat{\sigma}^i ( i \,\boldsymbol{\sigma}^i)\,.
\end{equation}
We find
\begin{equation}\begin{split}
		\hat{\sigma}^1&\equiv \sigma^1+\chi\,(-\cos(\alpha)\cos(\beta) \cos(\gamma)+\sin(\alpha)\sin(\gamma)){\rm d}\eta\,,\\
		\hat{\sigma}^2&\equiv \sigma^2-\chi\,(\sin(\alpha)\cos(\beta) \cos(\gamma)+\cos(\alpha)\sin(\gamma)){\rm d}\eta\,,\\
		\hat{\sigma}^3&\equiv \sigma^3+\chi\,\cos(\gamma)\sin(\beta){\rm d}\eta\,.
\end{split}\end{equation}
As we will show, the $D=10$ background for $\chi\neq 0$ can be obtained from the $\chi= 0$ solution given in \cite{Guarino:2020gfe} through the replacement
\begin{equation}
	\sigma^i\,\rightarrow\,\,\hat{\sigma}^i\,.\label{replacement}
\end{equation}
However, it is important to emphasise that the local coordinate redefinition \eqref{eq:CoordChange} is not globally well-defined and therefore does not define a diffeomorphism, except for the case $\chi = \frac{2\pi}{T}$, as shown clearly in section \ref{s:CStructure}. Hence, $\chi$ amounts to a physical modulus of the $D=10$ solution with periodicity $\frac{2\pi}T$.

\subsubsection{$\chi$ as a Complex Structure Modulus}\label{s:CStructure}
The parameter $\chi$ can also be interpreted as a complex structure modulus on $M_4 \sim S^3 \times S^1$, which gives another perspective on its geometric role and most clearly elucidates its periodicity $\chi \in [0,\frac{2\pi}T)$. For this, it is best to view $S^3$ as the Hopf fibration, such that the Hopf fibre and $S^1$ combine into an elliptic fibration over $S^2$. As we will now show, $\chi$ forms part of the complex structure modulus of the $T^2$ fibre.

We begin by considering a different, yet equivalent, parametrization of $S^3$ with coordinates ($\uphi$, $\xi$, $\psi$)\footnote{Their ranges are the same as the ($\alpha$, $\beta$, $\gamma$) ones.}, defined by
\begin{equation}
	\begin{split}
  g(\alpha, \beta, \gamma)& =g\left(0,\frac{\pi}{2},\pi\right)\cdot g(\uphi, \xi, \psi) \,.
  \end{split}
\end{equation}
A point in the four-dimensional total space that we are considering is now given by
\begin{equation}
 p = \left( \hat{g}(\uphi, \xi, \psi,\eta), \eta \right) \,,
\end{equation}
with
\begin{equation}
	\hat{g}(\uphi, \xi, \psi,\eta) \equiv h(\eta) \cdot g\left(0,\frac{\pi}{2},\pi\right) \cdot g(\uphi, \xi, \psi) \in \SU(2) \,.
\end{equation}
The projection map $\pi: M_4 \rightarrow S^2$ is essentially given by the usual Hopf map
\begin{eqnarray}
	\label{hopfmap}
	\pi : (\zeta_1, \zeta_2) \mapsto {\bf r} = \left({\rm Re}(2\zeta_1 \bar{\zeta_2}),\, {\rm Im}(2\zeta_1 \bar{\zeta_2}),\, |\zeta_1|^2-|\zeta_2|^2\right) \,,
\end{eqnarray}
with
\begin{equation}
	\begin{pmatrix}\zeta_1 & \zeta_2 \cr -\bar{\zeta}_2 & \bar{\zeta}_1\end{pmatrix} = g\left(0,\frac{\pi}{2},\pi\right)^{-1} \cdot \hat{g}(\uphi, \xi, \psi,\eta) \,.
\end{equation}
It is straightforward to check that ${\bf r}$ defined by \eqref{hopfmap} satisfies ${\bf r} \in S^2 \subset \mathbb{R}^3$ since ${\bf r} \cdot {\bf r} = 1$ and that $\psi$ and $\eta$ are projected out in \eqref{hopfmap}. Thus, $\psi$ and $\eta$ provide local coordinates on the $T^2$ fibre.

We can now read off the complex structure on the elliptic fibre, for example by studying the connection 1-forms on $M_4$. These are given by the right-invariant 1-forms
\begin{equation}
	\begin{split}
		\omega_{\psi} &= d\psi - 2 \chi d\eta + \cos\xi\, d\uphi \,, \\
		\omega_\eta &= d\eta \,.
	\end{split}
\end{equation}
Thus, the local holomorphic coordinate on the elliptic fibre is given by
\begin{equation}
	u = \psi + \hat{\tau}\, \eta \,,
\end{equation}
with $\hat{\tau} = i - 2 \chi$ defining the complex structure and the periodicity of $u$ given by
\begin{equation}
	u \sim u + 4\pi \sim u +\hat{\tau}\, T \,.
\end{equation}
Moreover, the $\hat{\sigma}^i$ now read
\begin{equation}
	\begin{split}
  \label{fibersigma}
  \hat{\sigma}^1 &= \frac{1}{2} (\sin (\xi ) \cos (\uphi ) (d\psi - 2\chi d\eta)- d\xi  \sin (\uphi )) \,, \\
  \hat{\sigma}^2 &= \frac{1}{2} (\sin (\xi ) \sin (\uphi ) (d\psi - 2\chi d\eta)+ d\xi \cos (\uphi )) \,, \\
  \hat{\sigma}^3 &= \frac{1}{2} (\cos (\xi ) (d\psi - 2\chi d\eta)+d\uphi ) \,,
  \end{split}
\end{equation}
or in terms of the complex coordinate $u$
\begin{equation}
	\begin{split}
		\label{fibersigmacomplex}
		\hat{\sigma}^1 &= \frac{1}{4} (\sin (\xi ) \cos (\uphi ) (du + d\bar{u})- 2 d\xi  \sin (\uphi )) \,, \\
		\hat{\sigma}^2 &= \frac{1}{4} (\sin (\xi ) \sin (\uphi ) (du + d\bar{u}) + 2 d\xi \cos (\uphi )) \,, \\
		\hat{\sigma}^3 &= \frac{1}{4} (\cos (\xi ) (du+d\bar{u})+d\uphi ) \,,
	\end{split}
\end{equation}
where there is no explicit dependence on $\chi$. Thus, it is clear that $\chi$ only affects the complex structure of the $T^2$ fibre.

The complex structure $\hat{\tau} = i - 2\chi$ now makes the periodicity of $\chi$ clear. First, recall that $\psi$ has periodicity $4\pi$ whereas $\eta$ has periodicity $T$. Let us thus rescale $\psi \rightarrow \psi' = \frac{\psi}{4\pi}$ and $\eta \rightarrow \eta' = \frac{\eta}{T}$ which have standard periodicities
\begin{equation}
	\psi' \sim \psi' + 1 \,, \qquad \eta' \sim \eta' + 1 \,.
\end{equation}
The local holomorphic coordinate, $u$, is given in terms of these by
\begin{equation}
	u = 4\pi \left( \psi' + \tau\, \eta' \right) \,,
\end{equation}
with the complex structure
\begin{equation}
	\tau = \frac{i}{4\pi} - \frac{\chi\,T}{2\pi} \,.
\end{equation}
It is now clear that $\chi\rightarrow \chi+\frac{2\pi}{T}$ just corresponds to a Dehn twist, $\tau \rightarrow \tau - 1$, and can be reabsorbed by a globally well-defined reparametrization. Thus, $\chi$ has periodicity $\frac{2\pi}{T}$.

\subsection{The Metric} \label{s:Metric}
The spacetime metric has the following form
\begin{equation}
ds^2=\frac{1}{2}\,\Delta^{-1}\,\left(ds^2_{{\rm AdS}_4}+ds^2_6\right)\,,
\end{equation}
where
\begin{equation}
\Delta\equiv (6-2\cos (2 \theta ))^{-\frac{1}{4}}\,.
\end{equation}
The internal metric $ds^2_6$ has the following form
\begin{equation}
  \label{twistedmetric}
ds^2_6=ds^2_{S^2}+ds^2_{S^3\times S^1}\,,
\end{equation}
where
\begin{equation}
ds^2_{S^2}=d\theta^2+\sin^2(\theta)\,d\varphi^2\,\,,\,\,\,ds^2_{S^3\times S^1}=\cos^2(\theta)\left(\hat{\sigma}_2^2+8\,\Delta^{4}\,(\hat{\sigma}_1^2+\hat{\sigma}_3^2)\right)+d\eta^2\,,
\end{equation}
and, for a fixed $\theta$, $S^3\times S^1$ denotes the twisted product described in the previous section. Note that the squashing of the $S^3$, arising from the different factors multiplying the $\hat{\sigma}^i$, breaks the $\SU(2) \times \SU(2)'$ symmetry of the round $S^3$ to $\SU(2) \times \U(1)'$, with the $\U(1)'$ rotating $\hat{\sigma}^1$ with $\hat{\sigma}^3$. As discussed in section \ref{GOTIS}, when $\chi \neq 0$, the $\SU(2)$ is also broken to $\U(1)$. Finally, the symmetries of $S^2$ are broken by the dependence on $\theta$ and $\varphi$ of the solution.

\subsection{The 2-Forms, the 4-Form, the Dilaton and the Axion}
As mentioned earlier, the expressions of the 2-forms and the 4-forms are the same as in the $\chi=0$ case, given in \cite{Guarino:2020gfe}, aside from the replacement $\sigma^i \rightarrow \hat{\sigma}^i$ as in \eqref{replacement}.

Thus, in the notation of \cite{Guarino:2020gfe}, we can write,
\begin{equation}
{B}^\alpha_{(2)}=A(\eta)^\alpha{}_\beta\, \mathfrak{b}^\beta_{(2)}\,,
\end{equation}
where
\begin{equation}
A(\eta)^\alpha{}_\beta\equiv \left(\begin{matrix}\cosh(\eta)& \sinh(\eta)\cr  \sinh(\eta) & \cosh(\eta)\end{matrix}\right)\,,
\end{equation}
is an ${\rm SL}(2,\mathbb{R})_{{\rm IIB}}$ twist and
\begin{equation}
	\begin{split}
  \mathfrak{b}^1_{(2)} &=\frac{1}{\sqrt{2}} \cos{(\theta)} \left[\left(\cos{(\phi)} \, d\theta + \frac{1}{2}\sin{(2\theta)} \, d(\cos{(\phi)})\right) \wedge \hat{\sigma}_2 + \cos{(\phi)} \frac{4 \sin(2 \theta)}{6-2 \cos(2\theta)} \hat{\sigma}_1 \wedge \hat{\sigma}_3 \right] \,,\\
  \mathfrak{b}^2_{(2)} &= -\frac{1}{\sqrt{2}} \cos{(\theta)} \left[\left(\sin{(\phi)} \, d\theta + \frac{1}{2}\sin{(2\theta)} \, d(\sin{(\phi)})\right) \wedge \hat{\sigma}_2 + \sin{(\phi)} \frac{4 \sin(2 \theta)}{6-2 \cos(2\theta)} \hat{\sigma}_1 \wedge \hat{\sigma}_3 \right]\,.
  \end{split}
\end{equation}
The self-dual 5-form field strength reads:
\begin{equation}
	\begin{split}
  \tilde{F}_5 &\equiv dC_{(4)}+\frac{1}{2}\epsilon_{\alpha\beta} B_{(2)}^\alpha\wedge H^\beta_{(3)}= (1 + \star )4 \Delta^4\sin{(\theta)} \cos^3{(\theta)} \left[ 3  \, d\theta \wedge d\phi \wedge \hat{\sigma}_1 \wedge \hat{\sigma}_2 \wedge \hat{\sigma}_3 \right.\\
  &\left.\qquad- d\eta \wedge \left(\cos(2 \theta) \, d\theta - \frac{1}{2} \sin{(2 \theta)} \sin{(2 \phi)} \, d\phi\right) \wedge \hat{\sigma}_1 \wedge \hat{\sigma}_2 \wedge \hat{\sigma}_3 \right]\,,
\end{split}
\end{equation}
where $H^\beta_{(3)}=dB_{(2)}^\alpha$.

Finally, the axion and the dilaton fields are encoded in the matrix $m_{\alpha\beta}$ in \eqref{mab} which, in our solution, reads
\begin{equation}
m_{\alpha\beta}=\AI^\sigma{}_\alpha\,\AI^\gamma{}_\beta\,\mathfrak{m}_{\sigma\gamma}\,,
\end{equation}
where
\begin{equation}
\mathfrak{m}_{\sigma\gamma}=2\,\Delta^2\,\left(
\begin{array}{cc}
 \sin ^2(\theta ) \cos ^2(\phi )+1 & -\frac{1}{2}
   \sin ^2(\theta ) \sin (2 \phi ) \\
 -\frac{1}{2} \sin ^2(\theta ) \sin (2 \phi ) & \sin
   ^2(\theta ) \sin ^2(\phi )+1 \\
\end{array}
\right)\,.
\end{equation}
Note that the axion-dilato system is the same as in the $\chi=0$ solution. This is due to the fact that the extra dependence on $\eta$ when $\chi\neq 0$ is entirely induced by the matrix $h(\eta)$ in \eqref{hatg}, and only affects those fields which depend on the point in $S^3$.

As mentioned earlier, the explicit dependence of the axion and dilaton on the coordinates $\theta,\,\varphi$ mean that the isometries of $S^2$ are not a symmetry of the whole solution, while ${\rm U}(1)^2={\rm U}(1) \times {\rm U}(1)'$ is. This is true for all values of $\chi$. For the special values of $\chi$
\begin{equation}
	\chi=\frac{m\pi}{T}\,, \qquad m \in \mathbb{Z} \,,
\end{equation}
the twist in the local product of $S^3\times S^1$ is either trivial ($m$ even) or $\mathbb{Z}_2$ ($m$ odd), as discussed at the end of subsection \ref{GOTIS}, and the symmetry of the solution is enhanced to ${\rm U}(2)={\rm SU}(2) \times {\rm U}(1)'$. Moreover, when $m$ is even, the solution is equivalent to $\chi = 0$, so that we should identify $\chi$ as a periodic modulus $\chi \sim \chi + \frac{2\pi}{T}$.

The dependence on $\eta$ through the ${\rm SL}(2,\mathbb{R})_{{\rm IIB}}$-twist matrix $A(\eta)^\alpha{}_\beta$ of the two-forms and the axio-dilaton system is the same as in the $\chi=0$ case, so we can apply to this family of solutions the same discussion about the corresponding ${\rm SL}(2,\mathbb{R})_{{\rm IIB}}$-monodromy matrix $\mathfrak{M}_{S^1}$ made in \cite{Guarino:2020gfe}. As $\eta\rightarrow \eta+T$ the twist matrix $A$ induces a monodromy
\begin{equation}
	\mathfrak{M}_{S^1}\equiv A^{-1}(\eta)\cdot A(\eta+T)=\begin{pmatrix}\cosh(T) & \sinh(T)\cr \sinh(T) & \cosh(T) \end{pmatrix}\,.
\end{equation}
By generalizing the twist matrix and suitably choosing the value of $T$ \cite{Guarino:2020gfe}, one can construct backgrounds in which the monodromy has the form
$\mathfrak{M}_{S^1}=-\mathcal{S}\,\mathcal{T}^k \in {\rm SL}(2,\mathbb{Z})_{{\rm IIB}}$, thus defining a family of S-fold solutions of Type IIB theory.

\subsection{The $\chi$-twist in the Kaluza-Klein spectrum}
As we have seen in section \ref{GOTIS}, a non-vanishing value of  $\chi$ induces an extra dependence on $\eta $ of those fields which, in the $\chi=0$ solution, were non-trivial functions of the point of $S^3$, due to the fibration of the latter over $S^1$. We can use this feature to determine the $\chi$-dependence on the full Kaluza-Klein tower of states. To do so, it is easiest to consider the background underformed, i.e.\ as for $\chi = 0$, and instead modify the Kaluza-Klein states' dependence on $\eta$. Thus, fields transforming in the $\SU(2)$ representation $\left[k\right]$, now acquire an $\eta$-dependence through the $\left[k \right]$-representation of the ${\rm SU}(2)$-element $h(\eta)$ given in \eqref{chitwist}. The corresponding twist matrix has eigenvalues
\begin{equation}
	e^{2\, i \,j\chi\,\eta}\,, \qquad \text{with } j=-k,\,-k+1,\dots,k-1,\, k\,.
\end{equation}

As an example, consider the three vector fields $A^i_\mu$, which, for $\chi=0$, gauge the ${\rm SU}(2) $ isometry group. For $\chi = 0$, these transform as the right-invariant Killing vectors $K_i$, defining the infinitesimal left-translations on $g(\alpha,\beta,\gamma)$. Indeed these vectors, on a group manifold, are defined as:
\begin{equation}
g^{-1}\cdot t_i\cdot g=K_i{}^\ell\sigma^s{}_\ell\,t_s\,,
\end{equation}
where $t_i$ are ${\rm SU}(2) $ generators, with $i = 1, 2, 3$, and we have written the left-invariant 1-forms $\sigma^i$ as $\sigma^i=\sigma^i{}_\ell\,dx^\ell$, $x^i\equiv(\alpha,\beta,\gamma)$.

When $\chi \neq 0$, the vector fields are modified. Transforming $g$ by the twist:
\begin{equation}
	g(\alpha,\beta,\gamma)\rightarrow \hat{g}(\alpha,\beta,\gamma,\eta)=h(\eta)\cdot g(\alpha,\beta,\gamma)\,,
\end{equation}
where $h(\eta)$ is the $2\times 2 $ twist matrix given in \eqref{chitwist}, we find
\begin{equation}
\hat{g}^{-1}\cdot t_i\cdot \hat{g}=g^{-1}\cdot h^{-1}\cdot t_i\cdot h \cdot g=h_i{}^\ell\,(K_\ell{}^k\sigma^s{}_k\,t_s)=\hat{K}_i{}^\ell\sigma^s{}_\ell\,t_s\,.
\end{equation}
Here $h_i{}^j$ denotes the adjoint action of $h$:
\begin{equation}
	h^{-1} \cdot t_i \cdot h \equiv h_i{}^j t_j \,.
\end{equation}
Therefore, as expected, the Killing vectors, and therefore $A^i_\mu$, transform in the $k=1$ representation acted on by the $3\times 3$ matrix $h^i{}_\ell(\eta)$.
The twisted vectors $\hat{A}_{(0)}{}^i_\mu$, at KK level $n=0$ on $S^1$, therefore now have a $\eta$-dependence due to the twist
\begin{equation}
\hat{A}_{(0)}{}^i_\mu(x,\eta)=h^i{}_\ell(\eta)\,A^\ell_\mu(x)\,.
\end{equation}
This additional $\eta$ dependence makes two of the vectors massive. As a result, $\SU(2)$ is broken at level $n=0$. Similarly, the corresponding vectors at level $n$ on $S^1$ have an $\eta$-dependence of the form
 \begin{equation}
\hat{A}_{(n)}{}^i_\mu(x,\eta)=h^i{}_\ell(\eta)\,A^\ell_\mu(x)e^{\frac{2 i \pi n\,\eta}{T}}\,, \qquad \hat{A}_{(n)}{}^i_\mu(x,\eta)^*=h^i{}_\ell(\eta)\,A^\ell_\mu(x)e^{-\frac{2 i \pi n\,\eta}{T}}\,.
\end{equation}

We can now see that when $\chi = p\pi/T$, with $p \in \mathbb{Z}$, the $\SU(2)$ symmetry is restored. Two of the eigenvalues of $\partial/\partial \eta$ on these vectors are now
\begin{equation}
	\pm \left(2  i \,\chi-\frac{2 i \pi n}{T}\right)=\pm \left(\frac{2  i \,\pi\,p}{T}-\frac{2 i \pi n}{T}\right)\,,
\end{equation}
which vanish for $n= p$. These correspond to the two gauge vectors at level $n > 0$ which become massless for these values of $\chi$ and enhance the ${\rm U}(1)$, seen at $S^1$ KK level $n=0$, back to ${\rm SU}(2)$. This is a bosonic version of the \emph{space invaders} scenario \cite{Duff:1986hr,Cesaro:2020piw}, where higher Kaluza-Klein modes become massless.

In general, on a field $\Phi_{(n)}^{[k ]}$, in $S^1$ KK level $n$ and in the $\left[k \right]$-representation of ${\rm SU}(2) $, the operator $\partial/\partial \eta$ will have eigenvalues
\begin{equation} \label{eq:ChiDependence}
\pm 2  i \,\left(j\,\chi-\frac{\pi n}{T}\right)\,,\,\,\,j=-k,\,-k+1,\dots,\,k-1,\, k\,,
\end{equation}
where $j$ can easily be identified with the $\U(1) \subset \SU(2)$ charge. The same conclusion can be reached by thinking of $\chi$ as part of the complex structure modulus of an elliptic fibration over $S^2$, as in section \ref{s:CStructure}. Now a field obtains an additional $\eta$-dependence by the replacement of the Hopf fibre coordinate $\psi \rightarrow \psi - 2\chi \eta$. As above, the $\psi$-dependence is determined by the field's $\U(1) \subset \SU(2)$ charge, so that the field's eigenvalues under $\partial/\partial \eta$ are again given by \eqref{eq:ChiDependence}. This explains the dependence on $\chi$ of the KK spectrum, as noted in section \ref{KKExFT}, see eq. \eqref{mmnj}.
\subsubsection{Summarizing the Symmetries in $\chi$ and Their Geometrical Interpretation}\label{finalchi}
Let us now summarize the understanding, which we have gained from the geometric description of the internal manifold, of the symmetries (\ref{sym1}), (\ref{sym2}) of the Kaluza-Klein spectrum. The former amounts to a Dehn twist of the toroidal fiber over $S^2$ which can be undone by a globally well defined reparametrization of the fiber. In particular for $\chi=2\pi/T$ the elliptic fibration is globally $S^3\times S^1$ where $S^3$ denotes the deformed three-sphere with isometry ${\rm SU}(2)\times {\rm U}(1)$, and thus the ${\rm U}(1)^2$ symmetry is enhanced to ${\rm SU}(2)\times {\rm U}(1)$.\par
As far as (\ref{sym2}) is concerned, the transformation $\chi\rightarrow -\chi$ corresponds to a transformation $\tau\rightarrow -\bar{\tau}$ in of the complex structure modulus of the toroidal fiber. This amounts in turn to a reflection in the imaginary axis of the torus, seen as a complex manifold, since it implies $u\rightarrow -\bar{u}$ as we also transform  $\psi\rightarrow -\psi$. It is not an invariance of the complex manifold itself, since it changes its orientation, but rather a parity transformation with respect to which the higher dimensional theory is invariant.
\footnote{For a discussion of parity symmetry in extended supergravities see \cite{Trigiante:2016mnt}.} Note that a change $\psi\rightarrow -\psi$ amounts, in the Kaluza-Klein modes, to changing the sign of the corresponding $j$ quantum number, as in (\ref{sym2}).

\subsection{More general ${\cal N}=2$ AdS$_4$}
The analysis of the Kaluza-Klein spectrum and of the internal space also suggests that a larger class of ${\cal N}=2$ AdS$_4$ S-fold vacua, which share many features with the family studied in the previous sections, can be obtained by performing a quotient on the latter solutions. However, we do not expect these new backgrounds to be vacua of the ${\cal N}=8$ supergravity.\par
Recall that the family of ${\cal N}=2$ AdS$_4$ vacua analysed here have an internal space, that is locally of the form $S^3 \times S^2 \times S^1$, with the $S^3$ non-trivially fibred over $S^1$ when $\chi \neq 0$. We can obtain ${\cal N}=2$ AdS$_4$ vacua with similar properties by replacing the $S^3$ by the Lens space $S^3/\mathbb{Z}_k$, with $k \in \mathbb{Z}^+$.

Since the quotient does not break the $\U(1)_R$ R-symmetry, the resulting AdS$_4$ vacua are still ${\cal N}=2$ supersymmetric. On the other hand, the $\mathbb{Z}_k$ quotient projects out various states, thus reducing the KK spectrum and isometries. In particular, at Kaluza-Klein level 0, only the states corresponding to 4-dimensional ${\cal N}=4$ supergravity survive the projection. This includes the modulus $\chi$. For $k = 2$, the vacua seem to admit a consistent truncation with 6 vector multiplets, while for $k \geq 3$, the vacua seem to admit a consistent truncation with 4 vector multiplets. These truncations can, in principle, be constructed by performing the $\mathbb{Z}_k$ quotient on the twist matrices \eqref{dressed_U} and assembling the invariant objects into a half-maximal structure \cite{Malek:2017njj}.

Finally, the $\mathbb{Z}_k$ quotient, for $k \geq 3$, breaks the isometries of the background, for all $\chi$, to $\U(1) \times \U(1)'$, while the $\mathbb{Z}_2$ quotient preserves the $\SU(2) \times \U(1)'$ isometry at $\chi = 0$. The periodicity of the modulus $\chi$ is also affected by the quotient. For a $\mathbb{Z}_k$ quotient, it is now given by $\chi \sim \chi + \frac{2\pi}{kT}$. This can be seen by noticing that for this value of $\chi$, the monodromy matrix $h(T) \in \mathbb{Z}_k$ now acts trivially on $S^3 / \mathbb{Z}_k$. The same conclusion can be reached by looking at the complex structure of the $T^2$ fibration over $S^2$ as in \ref{s:CStructure}, where the Hopf fibre of $S^3/\mathbb{Z}_k$ and the $S^1$ parametrized by $\eta$ make up the $T^2$ fibre. Since the Hopf fibre now has periodicity $\frac{4\pi}{k}$, the shift $\chi \rightarrow \chi + \frac{2\pi}{kT}$ corresponds to a Dehn twist.

\section{Discussion}
In this paper, we studied the one-parameter family of ${\cal N}=2$ AdS$_4$ vacua of $[{\rm SO}(6)\times {\rm SO}(1,1)]\ltimes \mathbb{R}^{12}$ gauged supergravity \cite{Guarino:2020gfe}. These uplift to 10-dimensional S-fold vacua of IIB supergravity of local form AdS$_4 \times S^5 \times S^1$. As the $S^1$ circle is traversed, the background undergoes an $\mathrm{SL}(2,\mathbb{Z})$ S-duality transformation. From the 4-dimensional perspective, the real parameter $\chi$, labeling the AdS$_4$ vacua, is non-compact with only one special point, $\chi = 0$, corresponding to a vacuum with $\SU(2) \times \U(1)_R$ symmetry, while all other vacua with $\chi \neq 0$ only have $\U(1) \times \U(1)_R$ symmetry in four dimensions.

We show that these features are in fact just four-dimensional artefacts and misrepresent the global properties of the conformal manifold of the dual CFT. In 10 dimensions, the parameter $\chi$ instead has periodicity equal to the inverse $S^1$ radius, $\frac{2\pi}{T}$. By computing the full Kaluza-Klein spectrum of the ${\cal N}=2$ AdS$_4$ vacua as a function of $\chi$, we showed explicitly that as $\chi \rightarrow \chi + \frac{2\pi}{T}$ the whole spectrum gets mapped to itself. However, what used to be modes of the 4-dimensional gauged supergravity now appear in the higher KK modes with non-zero $S^1$ level. This follows because the modulus $\chi$ remarkably only appears in the formula for the conformal dimension \eqref{Delta} through the combination
\begin{equation}
	\left(\frac{\pi n}{T} - j\,\chi\right)^2 \,,
\end{equation}
for a field of $\U(1) \subset \SU(2)$ charge $j$ and $S^1$ KK level $n$.

Another interesting characteristic of the Kaluza-Klein spectrum arises when $\chi = \frac{\pi}{T}$ the AdS$_4$ vacuum again has enhanced $\SU(2) \times \U(1)_R$ symmetry in 10 dimensions. The extra massless vector fields again come from higher KK modes, in a bosonic analogue of the \emph{space invader} scenario \cite{Duff:1986hr}. However, this vacuum is truly distinct from that corresponding to $\chi = 0$, with their Kaluza-Klein spectra differing. In particular, equation \eqref{Delta} shows that the masses of KK states of integer $\SU(2)$ spin are left invariant under $\chi \rightarrow \chi + \frac{\pi}{T}$, while those of half-integer $\SU(2)$ spin change under this shift.

Using the consistent truncation of \cite{Inverso:2016eet} to uplift the family ${\cal N}=2$ AdS$_4$ vacua to 10 dimensions, we were able to explain these features geometrically. In 10 dimensions, the modulus $\chi$ appears as a local coordinate transformation, which fails to be globally well-defined unless $\chi = \frac{2\pi}{T}$. In particular, $\chi$ induces a fibration of $S^3 \subset S^5$ over the $S^1$ with monodromy
\begin{equation}
	h(T) = \begin{pmatrix}
		\cos(\chi\,T) & \sin(\chi\,T) \\ -\sin(\chi\,T) & \cos(\chi\,T)
	\end{pmatrix} \,,
\end{equation}
making explicit the periodicity of $\chi \in [0,\frac{2\pi}T)$. The monodromy $h(T)$ breaks the $\SU(2)$ isometry to its commutant with $h(T)$ which, for generic values of $\chi$, is given by $\U(1)$. However, when $\chi = \frac{\pi}{kT}$, with $k \in \mathbb{Z}$, $h(T) \subset \mathbb{Z}_k$. This explains the special features observed when $\chi = \frac{\pi}{T}$, since now $h(T) = -{\bf 1}$, preserves all of $\SU(2)$ and leaves invariant all $\SU(2)$ integer-spin states.

Moreover, we show that the background remains invariant under the shift of the parameter $\chi \rightarrow \chi + \frac{2\pi}{T}$. This can be seen in a particularly clear way by writing the background as a $T^2$-fibration over $S^2$, which is further warped over $S^2$. As we showed, the parameter $\chi$ then appears as a complex structure modulus of the $T^2$-fibre and a shift $\chi \rightarrow \chi + \frac{2\pi}{T}$ corresponds to a Dehn twist of the $T^2$.

Another remarkable attribute of the Kaluza-Klein spectrum of the ${\cal N}=2$ S-fold backgrounds is that, for generic values of $\chi$, they consist only of long multiplets. We argue in section \ref{s:Spectrum} that this is due to the fact that the compactification includes a $S^1$ with tunable radius. Since the conformal dimension for states with non-zero Kaluza-Klein modes on $S^1$ must depend continuously on this tunable radius, these states must necessarily assemble into long multiplets\footnote{For special values of the radius, the conformal dimension of these states may hit the unitary bound, in which case these long multiplets will split into short multiplets.}. This general observation is quite powerful when computing the Kaluza-Klein spectrum. For example, when applied to the ${\cal N}=4$ AdS$_4$ vacuum of the same $[{\rm SO}(6)\times {\rm SO}(1,1)]\ltimes \mathbb{R}^{12}$ gauged supergravity \cite{Guarino:2020gfe}, this argument allows us to determine the full Kaluza-Klein spectrum from just the spin-2 spectrum, which has been worked out in \cite{Dimmitt:2019qla}. We give the result in appendix \ref{N4}.
A similar structure has been observed in the Kaluza-Klein spectrum on
{AdS}$_3\times {S}^3 \times {S}^3 \times {S}^1$ where the entire spectrum organises into long multiplets~\cite{deBoer:1999gea,Eberhardt:2017fsi,Baggio:2017kza,Eloy:2020uix} which can again be attributed
to the presence of the $S^1$ factor in the background.

Our work opens up several new questions. Firstly, it would be interesting to investigate other moduli of the ${\cal N}=2$ AdS$_4$ vacua which cannot be seen in 4-dimensional supergravity. For example, one may speculate that another modulus of the 10-dimensional AdS$_4$ vacua could be obtained by considering a general complex structure on the elliptic fibration used to describe the $S^3 \times S^1$ part of the internal space in section \ref{s:CStructure}. More generally, the infinitesimal moduli of the AdS$_4$ vacua can be constructed explicitly using our Kaluza-Klein fluctuation Ansatz \eqref{eq:FluctAnsatz}. However, to find their finite form, it may be worthwhile to use the techniques developed in \cite{Ashmore:2016oug,Ashmore:2018npi}.

The $[{\rm SO}(6)\times {\rm SO}(1,1)]\ltimes \mathbb{R}^{12}$ gauged supergravity also contains a symmetry-breaking family of ${\cal N}=1$ AdS$_4$ vacua, which can be analysed in a similar fashion to the study of ${\cal N}=2$ vacua presented here. This family of AdS$_4$ vacua is parametrized by two real scalar fields and generically has $\U(1) \times \U(1)$ symmetry. For special values of the real scalar fields, the symmetry is enhanced to $\SU(2) \times \U(1)$  or even $\SU(3)$. A natural question following from our work is if the true 10-dimensional moduli space of these ${\cal N}=1$ AdS$_4$ vacua is also compact and additional vacua have enhanced symmetries in 10-dimensions, with higher Kaluza-Klein vector fields becoming massless.

\section*{Acknowledgements}
We are grateful to P. Fr\'e, A. Guarino and C. Sterckx for helpful discussions. EM is supported by the Deutsche Forschungsgemeinschaft (DFG, German Research Foundation) via the Emmy Noether program ``Exploring the landscape of string theory flux vacua using exceptional field theory'' (project number 426510644).

\section*{Appendix}

\begin{appendix}
  \section{Superconformal multiplets}
  \label{supermult}
In this appendix, we summarize the structure of the relevant  ${\rm OSp}(2|4)$ multiplets,
and the pattern of multiplet shortening at critical values of the conformal dimensions,
as worked out in \cite{Minwalla:1997ka,Bhattacharya:2008zy}. We mostly follow the notation of~\cite{Cordova:2016emh},
to which we refer for details.

In short, the ${\rm OSp}(2|4)$ multiplet will be classified by Dynkin labels of the maximal compact subgroup $\mathrm{U}(1)\times \mathrm{SO}(3)_J \times \mathrm{U}(1)_\Delta$. The first factor represents the R--symmetry, whose charges we label by real $R \in \mathbb{R}$.
In accordance with the two independent supersymmetries $Q$ and $\overline{Q}$,
a generic ${\rm Osp}(2|4)$ multiplet will be of the form
\begin{equation}\begin{split}
X\bar{Y}[J]^{(R)}_\Delta\,,
\label{multipletXY}
\end{split}\end{equation}
where $X\in\{L,A_1,A_2,B_1\}$ and $\bar{Y}\in\{\bar{L},\bar{A}_1,\bar{A}_2,\bar{B}_1\}$
refer to the long and the different shortened structures with respect to $Q$ and $\overline{Q}$, respectively.
The parameters $R$, $J$, and $\Delta$ refer to the R--charge, Lorentz spin, and conformal dimension
of the highest weight state of the multiplet (\ref{multipletXY}), respectively.\footnote{
In contrast to the notation used in \cite{Cordova:2016emh}, our $J$ is half-integer,
referring to the spin, not the Dynkin label.
}

Unitarity implies the lower bound for the conformal dimension
\be
\Delta ~\ge~ 1+ |R| + J
\,.
\label{unitarityA}
\ee
For $\Delta > 1+ |R| + J$, the multiplet is of the long type $L\bar{L}[J]^{(R)}_\Delta$ and is given by the tensor product of its HWS with the representation
generated by the action of the 4 supercharges on a scalar vacuum, c.f.~section 4.2 of \cite{Cordova:2016emh}.
Therefore, its character factors as in (\ref{charactersLong}). Evaluating the product and
organizing the fields according to their Lorentz spins, yields the explicit field content which we
summarize in Tables~\ref{TabL0}, \ref{TabL1}, for $J=0, \frac12, 1$\,.
When the unitarity bound is saturated, the multiplets are shortened.
More precisely, for $R>0$ and $\Delta = 1+ R + J$, the right factor $\bar{L}$ in (\ref{multipletXY}) breaks according to
%For $\Delta > 1+ |R| + J$, the multiplet is of the long type $L\bar{L}[J]^{(R)}_\Delta$
%whose structure we summarize in Tables~\ref{TabL0}, \ref{TabL1} for $J=0, \frac12, 1$\,.
%When the unitarity bound is saturated, the multiplets are shortened.
%More precisely, for $R>0$ and $\Delta = 1+ R + J$, the right factor $\bar{L}$ in (\ref{multipletXY}) breaks according to
\begin{equation}\begin{split}
\bar{L}[1]_\Delta^{(R)} \longrightarrow
\bar{A}_1[1]_\Delta^{(R)} + \bar{A}_1[\tfrac12]_{\Delta+\frac12}^{(R+1)}
\,,\\
\bar{L}[\tfrac12]_\Delta^{(R)} \longrightarrow
\bar{A}_1[\tfrac12]_\Delta^{(R)} + \bar{A}_2[0]_{\Delta+\frac12}^{(R+1)}
\,,\\
\bar{L}[0]_\Delta^{(R)} \longrightarrow
\bar{A}_2[0]_\Delta^{(R)} + \bar{B}_1[\tfrac12]_{\Delta+1}^{(R+2)}
\,,
\label{shortening}
\end{split}\end{equation}
whereas for $R<0$ it is the left factor $L$ in (\ref{multipletXY}) which breaks accordingly.
At $R=0$, further shortening occurs, and massless multiplets show up.
We list the relevant shortened multiplets in Tables~\ref{TabS} and \ref{TabM},
where we restrict to the long-short case $L\bar{A}_1$, etc..
The short-long multiplets are obtained from the former upon replacing $R$ with $-R$ (we can refer to this operation as ''conjugation''). In particular, a half--hypermultiplet combined with its conjugate forms a complete hypermultiplet.

\begin{table}[h!]
 \begin{center}
\begin{tabular}{cc}
\begin{tabular}{|c|c|c|c|}
      \hline
      \multicolumn{4}{|c|}{${L\bar{L}[0]^{(R)}_{\Delta}}$:\;{long vector multiplet}} \\
      \hline
      spin & $\Delta$ & $R$ & $m^2$ \\ \hline
      \multirow{5}{*}{$0$} & $\Delta$ & $R$ & $(\Delta)(\Delta-3)$  \\
           & $\Delta+1$ & $R-2$ & $(\Delta+1)(\Delta-2)$  \\
           & $\Delta+1$ & $R+2$ & $(\Delta+1)(\Delta-2)$  \\
           & $\Delta+1$ & $R$ & $(\Delta+1)(\Delta-2)$  \\
           & $\Delta+2$ & $R$ & $(\Delta+2)(\Delta-1)$  \\
      \hline
      \multirow{4}{*}{$\frac{1}{2}$} & $\Delta+1/2$ & $R-1$ & $(\Delta-1)^2$ \\
           & $\Delta+1/2$ & $R+1$ & $(\Delta-1)^2$ \\
           & $\Delta+3/2$ & $R-1$ & $\Delta^2$ \\
           & $\Delta+3/2$ & $R+1$ & $\Delta^2$ \\
      \hline
      $1$ & $\Delta+1$ & $R$ & $(\Delta)(\Delta-1)$ \\ \hline
       \end{tabular}
&
     \begin{tabular}{|c|c|c|c|}
       \hline
       \multicolumn{4}{|c|}{${L\bar{L}[\textstyle\frac{1}{2}]^{(R)}_\Delta}$:\;{long gravitino multiplet}} \\
       \hline
       spin & $\Delta$ & $R$ & $m^2$ \\ \hline
       \multirow{4}{*}{$0$} & $\Delta+1/2$ & $R-1$ & $(\Delta+1/2)(\Delta-5/2)$  \\
            & $\Delta+1/2$ & $R+1$ & $(\Delta+1/2)(\Delta-5/2)$  \\
            & $\Delta+3/2$ & $R-1$ & $(\Delta+1/2)(\Delta-5/2)$  \\
            & $\Delta+3/2$ & $R+1$ & $(\Delta+3/2)(\Delta-3/2)$  \\
       \hline
       \multirow{6}{*}{$\frac{1}{2}$} & $\Delta$ & $R$ & $(\Delta-3/2)^2$ \\
            & $\Delta+1$  & $R-2$ & $(\Delta-1/2)^2$ \\
            & $\Delta+1$  & $R+2$ & $(\Delta-1/2)^2$ \\
            & $\Delta+1$  & $R$ & $(\Delta-1/2)^2$ \\
            & $\Delta+1$  & $R$ & $(\Delta-1/2)^2$ \\
            & $\Delta+2$  & $R$ & $(\Delta+1/2)^2$ \\ \hline
       \multirow{4}{*}{$1$} & $\Delta+1/2$ & $R-1$ & $(\Delta-1/2)(\Delta-3/2)$  \\
            & $\Delta+1/2$ & $R+1$ & $(\Delta-1/2)(\Delta-3/2)$  \\
            & $\Delta+3/2$ & $R-1$ & $(\Delta+1/2)(\Delta-1/2)$  \\
            & $\Delta+3/2$ & $R+1$ & $(\Delta+1/2)(\Delta-1/2)$  \\ \hline
       $3/2$ & $\Delta+1$ & $R$ & $(\Delta-1/2)^2$ \\ \hline
     \end{tabular}
\end{tabular}
\caption{Long ${\cal N}=2$ multiplets ${L\bar{L}[0]^{(R)}_{\Delta}}$ and ${L\bar{L}[\textstyle\frac{1}{2}]^{(R)}_\Delta}$\,.}
\label{TabL0}
\end{center}
\end{table}

 \begin{table}[h!]
 \begin{center}
 \begin{tabular}{|c|c|c|c|}
    \hline
    \multicolumn{4}{|c|}{${L\bar{L}[1]^{(R)}_\Delta}$:\;{long graviton multiplet}} \\
    \hline
    spin & $\Delta$ & $R$ & $m^2$ \\ \hline
    $0$ & $\Delta$ & R & $\Delta(\Delta-3)$  \\
    \hline
    \multirow{4}{*}{$\frac{1}{2}$} & $\Delta+1/2$ & $R-1$ & $(\Delta-1)^2$ \\
         & $\Delta+1/2$  & $R+1$ & $(\Delta-1)^2$ \\
         & $\Delta+3/2$  & $R-1$ & $\Delta^2$ \\
         & $\Delta+3/2$  & $R+1$ & $\Delta^2$ \\
    \hline
    \multirow{6}{*}{$1$} & $\Delta$ & $R$ & $(\Delta-1)(\Delta-2)$  \\
         & $\Delta+1$ &  $R$ & $\Delta(\Delta-1)$  \\
         & $\Delta+1$ & $R$ & $\Delta(\Delta-1)$  \\
         & $\Delta+1$ & $R-2$ & $\Delta(\Delta-1)$  \\
         & $\Delta+1$ & $R+2$ & $\Delta(\Delta-1)$  \\
         & $\Delta+2$ & $R$ & $\Delta(\Delta+1)$  \\
    \hline
    \multirow{4}{*}{$\frac{3}{2}$} & $\Delta+1/2$ & $R-1$ & $(\Delta-1)^2$  \\
         & $\Delta+1/2$ &  $R+1$ & $(\Delta-1)^2$  \\
         & $\Delta+3/2$ & $R-1$ & $\Delta^2$  \\
         & $\Delta+3/2$ & $R+1$ & $\Delta^2$  \\
    \hline
    $2$ & $\Delta+1$ & $R$ & $(\Delta+1)(\Delta-2)$ \\ \hline
  \end{tabular}
\caption{Long ${\cal N}=2$ multiplet ${L\bar{L}[1]^{(R)}_{\Delta}}$\,.}
\label{TabL1}
\end{center}
\end{table}

\begin{table}[h!]
 \begin{center}
\begin{tabular}{cc}
    \begin{tabular}{|c|c|c|c|}
      \hline
      \multicolumn{4}{|c|}{${L\bar{A}_1[\textstyle\frac{1}{2}]^{(R)}_{R+\frac{3}{2}}}$:\;{short masssive gravitino multiplet}} \\
      \hline
      spin & $\Delta$ & $R$ & $m^2$ \\ \hline
      $0$ & $R+2$ & $R-1$ & $(R+2)(R-1)$ \\ \hline
      \multirow{3}{*}{$\frac{1}{2}$} & $R+3/2$ & $R$ & $R^2$ \\
           & $R+5/2$ & $R-2$ & $(R+1)^2$ \\
           & $R+5/2$ & $R$ & $(R+1)^2$ \\ \hline
      \multirow{3}{*}{$1$} & $R+2$ & $R-1$ & $R(R+1)$ \\
           & $R+2$ & $R+1$ & $R(R+1)$ \\
           & $R+3$ & $R-1$ & (R+2)(R+1) \\ \hline
      $3/2$ & $R+5/2$ & $R$ & $(R+1)^2$ \\ \hline
    \end{tabular}
  &
     \begin{tabular}{|c|c|c|c|}
       \hline
       \multicolumn{4}{|c|}{${L\bar{B}_1[0]^{(R)}_{R}}$:\;{half-hypermultiplet}} \\
       \hline
       spin & $\Delta$ & $R$ & $m^2$ \\ \hline
       \multirow{2}{*}{$0$} & $R$ & $R$ & $R(R-3)$ \\
            & $R+1$ & $R-2$ & $(R+1)(R-2)$ \\ \hline
       $1/2$ & $R+1/2$ & $R-1$ & $(R-1)^2$ \\ \hline
     \end{tabular}
\end{tabular}
\caption{Shortened ${\cal N}=2$ multiplets.}
\label{TabS}
\end{center}
\end{table}

\begin{table}[h!]
\begin{center}
\begin{tabular}{cc}
    \begin{tabular}{|c|c|c|c|}
      \hline
      \multicolumn{4}{|c|}{${A_1\bar{A}_1[1]^{(0)}_2}$:\;{massless graviton multiplet}} \\
      \hline
      spin & $\Delta$ & $R$ & $m^2$ \\ \hline
      $1$ & $2$ & $0$ & $0$ \\ \hline
      \multirow{2}{*}{$\frac{3}{2}$} & $5/2$ & $-1$ & $1$ \\
           & $5/2$  & $+1$ & $1$ \\ \hline
      $2$ & $3$ & $0$ & $0$ \\ \hline
    \end{tabular}
  &
     \begin{tabular}{|c|c|c|c|}
       \hline
       \multicolumn{4}{|c|}{${A_2\bar{A}_2[0]^{(0)}_1}$:\;{massless vector multiplet}} \\
       \hline
       spin & $\Delta$ & $R$ & $m^2$ \\ \hline
       \multirow{2}{*}{$0$} & $1$ & $0$ & $-2$ \\
            & $2$ & $0$ & $-2$ \\ \hline
       \multirow{2}{*}{$\frac{1}{2}$} & $3/2$ & $-1$ & $0$ \\
            & $3/2$ & $+1$ & $0$ \\ \hline
       $1$ & $2$ & $0$ & $0$ \\ \hline
     \end{tabular}
\end{tabular}
\caption{Massless ${\cal N}=2$ multiplets.}
\label{TabM}
\end{center}
\end{table}

\section{Kaluza-Klein spectrum of the $\mathcal{N}=4$ Vacuum}\label{N4}

The $\mathcal{N}=4$ AdS$_4$ vacuum, first presented in \cite{Gallerati:2014xra}, is defined by the following expectation values of the $z_i$:
\begin{equation}
  z_1=z_2=z_3=i\,, \qquad z_4=z_5=z_6=-\overline{z}_7=\frac{1}{\sqrt{2}}(1+i) \,.
\end{equation}
The vacuum has $\SU(2) \times \SU(2)$ symmetry, corresponding to the superconformal R-symmetry.

As argued in section \ref{s:Spectrum}, the fact that the background contains a $S^1$, whose radius can be varied, implies that the Kaluza-Klein states with non-zero modes on the $S^1$ must fit into long supermultiplets. Moreover, by decomposing the ${\cal N}=8$ multiplet into long ${\cal N}=4$ multiplets, we deduce that, for generic values of the $S^1$ radius, the entire Kaluza-Klein spectrum organises itself into long \textit{graviton} multiplets. Therefore, the full Kalzua-Klein spectrum of the ${\cal N}=4$ vacuum can be determined from just its spin-2 spectrum, which has been worked out in \cite{Dimmitt:2019qla}.

Indeed, a direct computation using the tools of \cite{Malek:2019eaz,Malek:2020yue} and reviewed in section \ref{KKExFT} confirms that all Kaluza-Klein modes can be organised into long graviton multiplets. These are counted by the character for the highest-weight states, i.e. the gravitons,
\begin{equation} \label{eq:N4char}
	\nu_1 = \frac{1}{\left(1-q^2\right)\left(1-q\,u\right)\left(1-q\,v\right)} \frac{1+s}{1-s} \,.
\end{equation}
Here exponents of $q$, $s$ count levels for the $S^5$ and $S^1$ harmonics, respectively, while exponents of $u$, $v$ count the $\SU(2) \times \SU(2)$ spins.

We find that the conformal dimension, $\Delta$, of the highest weight state of the supermultiplets, as counted by \eqref{eq:N4char}, is given by
\begin{equation}
	\Delta = \frac32 + \frac12 \sqrt{9 + 2\ell (\ell+4) + 4\ell_1(\ell_1+1) + 4\ell_2(\ell_2+1) + \frac{2n^2\pi^2}{T^2}} \;,
\end{equation}
for a HWS of type $q^{\ell}\,s^n\,u^{\ell_1}\,v^{\ell_2}$. This precisely matches the spin-2 Kaluza-Klein masses computed in \cite{Dimmitt:2019qla}.
\end{appendix}

\providecommand{\href}[2]{#2}\begingroup\raggedright\endgroup

\end{document}